\documentclass{aa}  
\usepackage{graphicx}
\usepackage{txfonts}
\usepackage{xcolor}

\defcitealias{Fioretti2024}{Paper~I}
\DeclareUnicodeCharacter{2212}{-}
\begin{document}

   \title{Unveiling the origin of {\it XMM-Newton} soft proton flares}

   \subtitle{II. Systematics in the proton spectral analysis }

   \author{T. Mineo
          \inst{1}\fnmsep\thanks{\email{teresa.mineo@inaf.it}},         
          V. Fioretti
          \inst{2},
          S. Lotti
          \inst{3},
          S. Molendi
         \inst{4},
          G. Lanzuisi
          \inst{2},
         M. Cappi
        \inst{2},
         M. Dadina
        \inst{2},
         S. Ettori
        \inst{2},
         F. Gastaldello 
        \inst{4},
        R. Amato
        \inst{5}
          }

   \institute{INAF, Istituto di Astrofisica Spaziale e Fisica Cosmica (IASF) di Palermo, Via Ugo La Malfa 153, 90146 Palermo, Italy
   \and
   INAF, Osservatorio di Astrofisica e Scienza dello Spazio (OAS) di Bologna, via P. Gobetti 93/3, 40129 Bologna, Italy
         \and
        INAF, Istituto di Astrofisica e Planetologia Spaziali (IAPS), Via del Fosso del Cavaliere 100, 00133 Roma, Italy
       \and
       INAF, Istituto di Astrofisica Spaziale e Fisica Cosmica (IASF) di Milano, Via E. Bassini 15, 20133 Milano, Italy
       \and
       INAF, Osservatorio Astronomico di Roma (OAR), Via Frascati 33, 00078 Monte Porzio Catone (RM), Italy
             }

   \date{ }
 
  \abstract
{Low-energy ($<$ 300 keV) protons entering the field of view of the {\it XMM-Newton} telescope  scatter with the X-ray mirror surface and might reach the X-ray detectors on the focal plane. They manifest in the form of a sudden increase in the rates, usually referred to as soft proton flares. By knowing the conversion factor between the soft proton energy and the deposited charge on the detector, it is possible to derive the incoming flux and to study the environment of the Earth magnetosphere at different distances, given the wide and elliptical {\it XMM-Newton} orbit.  Thanks to detailed Geant4 simulations, we were able to build specific soft proton response matrices for MOS and PN.
}
{In this second paper, we present the results of testing these matrices with real data for the first time, while also exploring the seasonal and solar activity effect on the proton environment. The selected spectra are relative to 55 simultaneous MOS and PN observations with flares raised in four different temporal windows: December-January and July-August of 2001-2002 (solar maximum) and 2019-2020 (solar minimum).
}
{We selected and extracted the flare mean spectra and count rates in the 2 -- 11.5 keV energy range for the four epochs. After investigating the rate variations among the MOS1, MOS2, and PN instruments, we fit the X-ray spectra using XSPEC and the proton response matrices. The best-fitting parameters derived for the three instruments were compared in order to obtain the systematic errors.}
{There is no seasonal or solar activity effect on the soft proton mean count rates, but we find large discrepancies in the instrument cross-correlations across the 20 years of satellite operations. In 2001-2002, after a few years of operation, the MOS1 and MOS2 rates are similar, and about 20\% with regard to the PN ones. 
After 20 years, PN does not present any variation in its response, while MOS1 suffers a reduction of $\sim$30\%, in addition to the 30\% loss due to the damage of two CCDs, and MOS2 is affected by an even worse degradation (70\%). 
The main result of the spectral analysis is that the physical model representative of the proton spectra at the input of the telescope is a power law. However, a second and phenomenological component is necessary to take into account imprecision in the generation of the matrices at softer ($<$5keV) energies. This component contributes for 21\% for the MOS and 5\% for the PN to the total flux in the 2–5 keV energy range. }
{This study, which is the first application of the soft proton response matrices to real data, shows coherent results between detectors and allows us to estimate  systematic uncertainties in the measured spectra of 3\% between the two MOS detectors and of 24\% between MOS and PN, together with a systematic in the input flux of about a factor of two. They are all likely due to uncertainties in the proton transmission models, with the presence of additional passive material in front of the front-illuminated MOS, and element deposition on its electrode structure across the mission life. Dedicated studies and laboratory measurements are required for improving the accuracy of the proton response files.
}

\titlerunning{Unveiling the origin of {\it XMM-Newton} soft proton flares II}
\authorrunning{T. Mineo  et al}
\maketitle
%

\section{Introduction}
The X-ray telescope capability of focusing protons with energies lower than a
few hundred kiloelectron volts, so-called soft protons, was discovered just after the launch of the {\it Chandra} X-ray observatory \citep{Weisskopf2002}, when a rapid degradation of the front illuminated
CCDs at the focal plane of the Wolter type-1 telescope occurred \citep{Lo2003}. The damages were in fact caused by soft protons encountered along the orbit that, hitting the X-Ray optics surface at grazing angles, were transmitted to the focal plane. 

{\it XMM-Newton} \citep{Jansen2001}  has  been exposed to the same risk, carrying three Wolter type-1 mirrors to focus X-rays into the focal plane instrument EPIC (European Photon Imaging Cameras),
consisting of three cameras, covering the 0.2 -15 keV energy range: two MOS \citep{Turner2001} and one PN CCDs \citep{Struder2001}.  
The focal plane detectors are protected from the background induced by IR, visible, and UV light with three optical blocking filters with different thicknesses of both the aluminum film and its support: the thin, medium, and thick filters. They are mounted on a filter wheel and can be changed depending on the requirement on the observations. In addition, the two MOS telescopes are equipped with reflection grating spectrometer \citep[RGS,][]{denHerder2004} nested behind the mirrors.
These instruments, operated simultaneously with EPIC, allow for high resolution (E/$\Delta$E ranging from 100 to 500) measurements in the soft X-ray range of 0.35 -- 2.5 keV. 

The soft protons transmitted by the optics, after crossing the optical filters, release a part of their energy in the EPIC CCDs,
generating a local charge indistinguishable from the signal produced by photons. 
In the case of a sudden increase in the input flux, the final effect is the generation of flaring events, lasting from hundreds of seconds to hours, which degrade the quality of science observations because of the increased background, resulting in an average loss of 30--40\% of exposure \citep{Marelli2017}.

However, what is a disadvantage for the  observation of  astronomical sources would be a good opportunity to study the environment crossed by the satellite along its orbit, provided that a proton energy-transfer function is available. 
As is explained in detail in \citet[][hereafter Paper I]{Fioretti2024}, this function in the format of a response matrix has been built from detailed Geant4 simulations modeling all the proton interactions with the telescope elements: the transmission from optics, the energy degradation from the filters, and the total release of the soft proton energy in the detectors.
The RGS was not included in the simulator. However, considering that it shades $\sim$50\% of the photon flux in the EPIC/MOS \citep{denHerder2004}, the same attenuation factor was assumed for protons and the total effective area of this instrument was halved independently of the energy \citepalias{Fioretti2024}.
This assumption is confirmed by the simulations performed by \citet{Nartallo2002} to quantify the level of soft protons focused by the optics to {\it XMM-Newton} and {\it Chandra} focal planes. 

In summary, three matrices for the MOS and three for the PN, one for each filter, have been generated. They are relative to a single event pattern with no spatial selection at the focal plane.  The working energy range is 2--11.5 keV for the MOS and 2--19.5 for the PN, with a systematic uncertainty on the absolute flux of $\sim$20\%.  As is explained in \citetalias{Fioretti2024}, a first validation of the matrices has been performed using representative MOS spectra obtained from the EXTraS\footnote{http://www.extras-fp7.eu} (Exploring the X-ray Transient and variable Sky) archive of focused background spectra \citep{Salvetti2017}, accumulated in the first 13 years of {\it XMM-Newton} life.  

\begin{figure}[t]
\centering
\includegraphics[width=5.5cm, angle= -90]{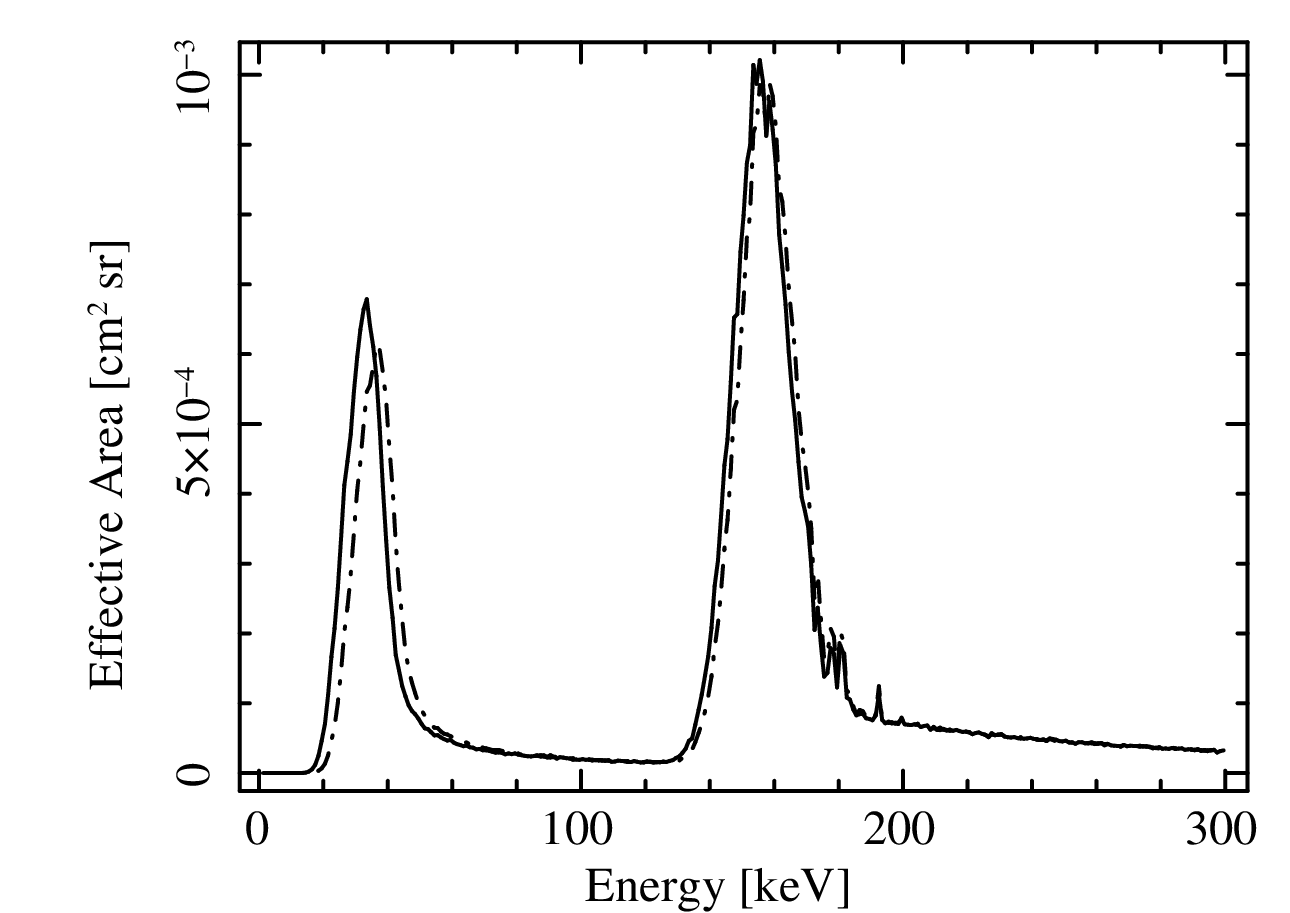}

\hspace{1.0cm}

\includegraphics[width=5.5cm, angle= -90]{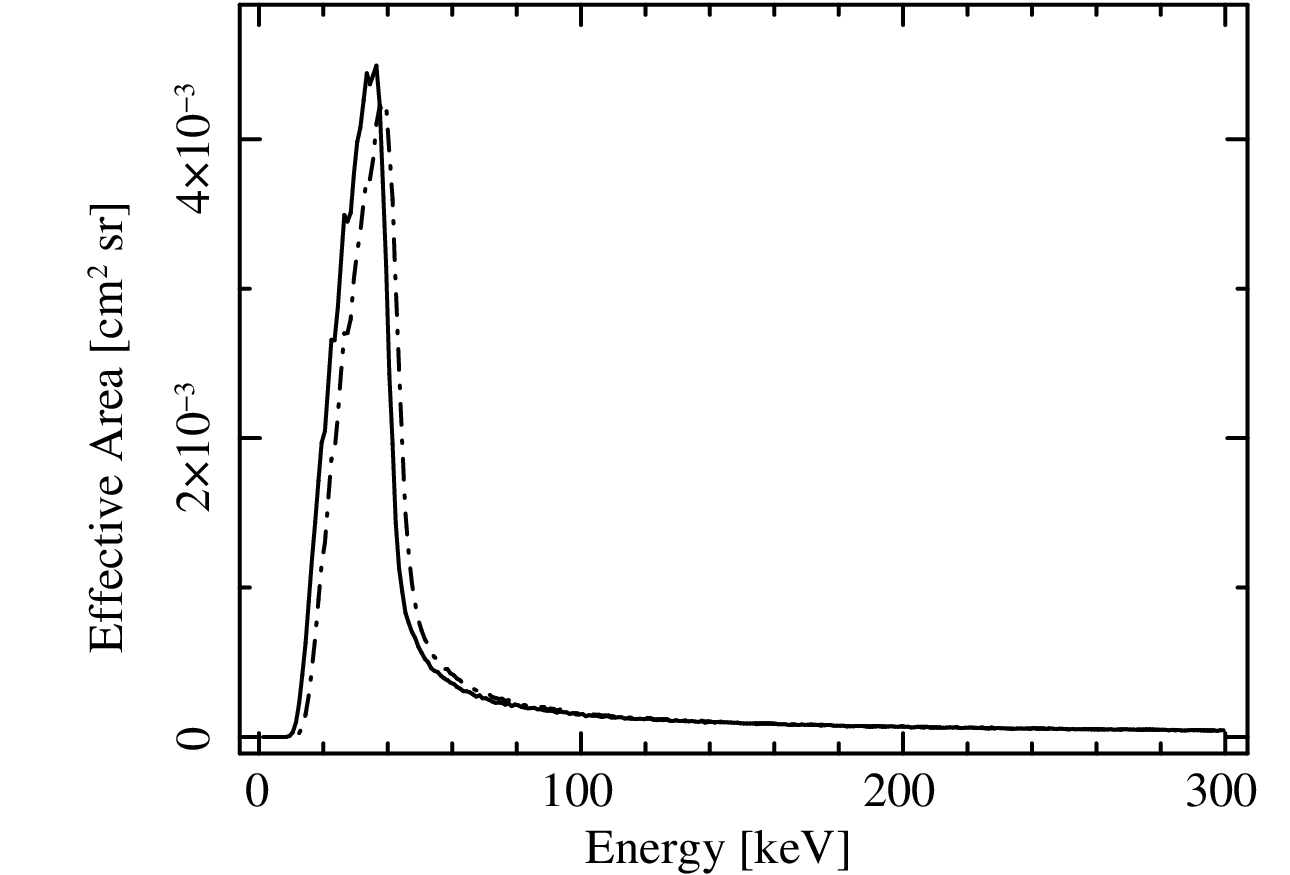}
\caption{Proton response matrix effective area for MOS (top panel) and PN (bottom panel). 
The application of the thin (medium) filter is represented by a continuous (dash-dotted) curve. }
\label{fig:effarea}
\end{figure}

The proton response matrices\footnote{The first release for the proton RMF and ARF files of {\it XMM-Newton} is publicly distributed in https://zenodo.org/records/7629674,
\\
while the latest releases and updates can be found in  
\\
https://www.ict.inaf.it/gitlab/proton\_response\_matrix}
have been produced to be compliant with XSPEC \citep{Dorman2001}, an interactive X-ray spectral-fitting program, and are given as two separate files:  {\it i)} the ancillary response file that stores the summed contributions of all efficiencies; and {\it ii)} the redistribution matrix that contains the probability that an incoming proton of energy, $E$, is detected in the output detector channel, $PHA$. The effective area relative to MOS and PN for the thin and medium filter are shown in Fig.~\ref{fig:effarea}. The two peaks in the MOS curves are generated by the complex structure of the open electrode with two regions of different thickness \citepalias{Fioretti2024}.

\begin{figure}[th]
\centering
\includegraphics[width=5.5cm, angle= -90]{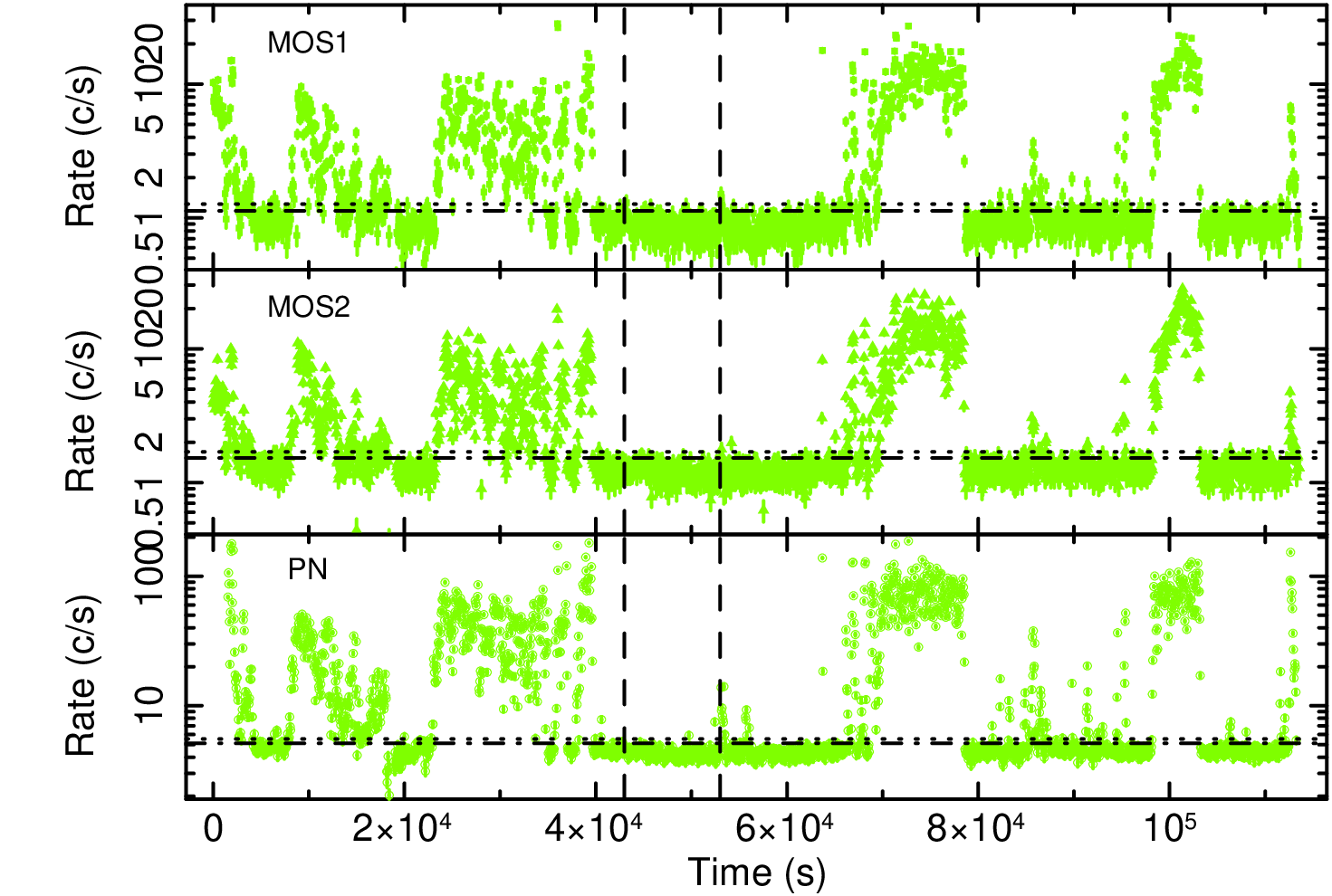}
\caption{Light curves relative to observation ID 0844860601 (2020-01-06) for MOS1, MOS2, and PN. 
The dashed vertical lines delimit the interval in which the background rate was evaluated, the dash-dotted lines indicate the highest background rate, and the dotted lines indicate the lowest flare rate.}
\label{fig:src_selection}
\end{figure}

\begin{table*} [th]
\caption{\label{tab:log_obs1} Log of the observations considered for matrix validation during the solar maximum activity.  }             
\begin{tabular}{c | c | c c c | c c c | c }     
\hline\hline       
 Observation ID&  \multicolumn{1}{c|} {Date}& \multicolumn{3}{c|} {Exposure} & \multicolumn{3}{c|} {2-11.5 keV Rate} & Filter \\
   & (y-m-d) &\multicolumn{3}{c|} {(s)} & \multicolumn{3}{c|} {(c/s)} & \\
   &   & MOS1 & MOS2 & PN  & MOS1 & MOS2 & PN  & \\
\hline   
0011830201 & 2001-12-08 & 7205.1 & 7265.6 & 4798.1 & 20.76 $\pm$ 0.05 & 20.78 $\pm$ 0.05 & 69.23 $\pm$ 0.12 & Thin1 \\
0027340101 & 2002-01-03 & 9541.0 & 9540.8 & 8522.4 & 3.59 $\pm$ 0.02 & 3.40 $\pm$ 0.02 & 23.55 $\pm$ 0.05 & Thin1 \\
0033540901 & 2002-01-06 & 2163.2 & 2162.4 & 1875.7 & 5.54 $\pm$ 0.05 & 5.31 $\pm$ 0.05 & 29.95 $\pm$ 0.13 & Thin1 \\
0037980401 & 2002-01-26 & 2234.0 & 2248.0 & 1894.0 & 10.48 $\pm$ 0.07 & 11.04 $\pm$ 0.07 & 40.28 $\pm$ 0.15 & Thin1 \\
0052140201 & 2001-12-02 & 6731.2 & 6734.9 & 5907.8 & 6.60 $\pm$ 0.03 & 5.86 $\pm$ 0.03 & 38.25 $\pm$ 0.08 & Medium \\
0071340201 & 2001-12-27 & 7920.9 & 7925.6 & 6909.0 & 2.42 $\pm$ 0.02 & 2.31 $\pm$ 0.02 & 11.91 $\pm$ 0.05 & Medium \\
0071340501 & 2001-12-24 & 17728.9 & 17738.6 & 15133.9 & 6.32 $\pm$ 0.02 & 6.27 $\pm$ 0.02 & 29.66 $\pm$ 0.05 & Thin1 \\
0085280301 & 2002-01-28 & 5075.3 & 5078.9 & 4574.7 & 3.52 $\pm$ 0.03 & 3.07 $\pm$ 0.03 & 23.07 $\pm$ 0.08 & Medium \\
0108060601 & 2002-01-13 & 15904.6 & 15845.5 & 13092.3 & 10.38 $\pm$ 0.03 & 10.07 $\pm$ 0.03 & 39.89 $\pm$ 0.06 & Thin1 \\
0108060701 & 2002-01-14 & 16797.8 & 16812.2 & 14804.0 & 2.13 $\pm$ 0.01 & 2.13 $\pm$ 0.01 & 10.62 $\pm$ 0.03 & Thin1 \\
0108061901 & 2002-01-17 & 8571.3 & 8614.9 & 5538.2 & 23.56 $\pm$ 0.05 & 22.09 $\pm$ 0.05 & 117.83 $\pm$ 0.15 & Thin1 \\
0108062301 & 2002-01-23 & 10003.4 & 10005.9 & 8956.2 & 1.00 $\pm$ 0.01 & 0.96 $\pm$ 0.01 & 5.44 $\pm$ 0.03 & Thin1 \\
0109270301 & 2002-01-26 & 27208.4 & 27246.6 & 24335.6 & 4.69 $\pm$ 0.01 & 4.56 $\pm$ 0.01 & 24.79 $\pm$ 0.03 & Medium \\
0112551101 & 2001-12-16 & 1686.7 & 1685.0 & 1412.2 & 5.51 $\pm$ 0.06 & 6.12 $\pm$ 0.06 & 17.82 $\pm$ 0.12 & Thin1\tablefootmark{*} \\
0112681001 & 2002-01-30 & 16873.7 & 16955.9 & 15387.7 & 2.82 $\pm$ 0.01 & 2.85 $\pm$ 0.01 & 16.12 $\pm$ 0.04 & Thin1 \\
&   &      &      &     &      &      &     & \\
0037981201 & 2002-07-15 & 1469.6 & 1470.4 & 1292.5 & 6.39 $\pm$ 0.07 & 6.16 $\pm$ 0.07 & 35.61 $\pm$ 0.17 & Thin1 \\
0037981601 & 2002-07-26 & 1068.9 & 1069.5 & 964.6 & 1.48 $\pm$ 0.04 & 1.28 $\pm$ 0.04 & 8.43 $\pm$ 0.10 & Thin1 \\
0037982701 & 2002-08-15 & 2058.7 & 2063.4 & 1811.4 & 1.02 $\pm$ 0.03 & 0.97 $\pm$ 0.03 & 5.33 $\pm$ 0.06 & Thin1 \\
0056022201 & 2002-08-17 & 12221.2 & 12283.9 & 10107.3 & 8.62 $\pm$ 0.03 & 8.66 $\pm$ 0.03 & 30.19 $\pm$ 0.06 & Thin1 \\
0093160201 & 2002-08-23 & 8665.4 & 8806.1 & 5429.9 & 21.76 $\pm$ 0.05 & 20.50 $\pm$ 0.05 & 78.29 $\pm$ 0.12 & Thin1 \\
0103060201 & 2002-08-01 & 14343.6 & 14351.6 & 12605.5 & 6.38 $\pm$ 0.02 & 6.08 $\pm$ 0.02 & 26.33 $\pm$ 0.05 & Thin1 \\
0106060501 & 2002-07-06 & 1891.0 & 1891.4 & 1356.5 & 22.58 $\pm$ 0.11 & 23.67 $\pm$ 0.11 & 80.85 $\pm$ 0.25 & Thin1 \\
0109460801 & 2002-08-10 & 6063.1 & 6065.2 & 5434.7 & 5.83 $\pm$ 0.03 & 5.85 $\pm$ 0.03 & 30.53 $\pm$ 0.08 & Thin1 \\
0109661201 & 2002-07-16 & 12360.2 & 12370.5 & 10785.2 & 6.21 $\pm$ 0.02 & 6.15 $\pm$ 0.02 & 31.05 $\pm$ 0.06 & Thin1 \\
0110980601 & 2002-07-05 & 23456.4 & 23483.2 & 20154.8 & 6.45 $\pm$ 0.02 & 6.38 $\pm$ 0.02 & 32.86 $\pm$ 0.04 & Thin1 \\
0111281501 & 2002-07-28 & 2007.0 & 2005.3 & 1783.0 & 6.95 $\pm$ 0.06 & 6.71 $\pm$ 0.06 & 38.86 $\pm$ 0.15 & Thin1 \\
0111281601 & 2002-07-20 & 311.7 & 311.7 & 276.4 & 7.61 $\pm$ 0.16 & 7.81 $\pm$ 0.16 & 31.71 $\pm$ 0.35 & Thin1 \\
0111290601 & 2002-07-28 & 20892.2 & 20956.0 & 15798.4 & 12.20 $\pm$ 0.03 & 11.67 $\pm$ 0.02 & 52.97 $\pm$ 0.06 & Thin1 \\
0112370601 & 2002-08-12 & 12995.5 & 13002.3 & 11296.5 & 5.46 $\pm$ 0.02 & 5.36 $\pm$ 0.02 & 26.05 $\pm$ 0.05 & Thin1 \\
0112370801 & 2002-08-09 & 11345.9 & 11360.5 & 9044.0 & 10.90 $\pm$ 0.03 & 10.52 $\pm$ 0.03 & 45.49 $\pm$ 0.07 & Thin1 \\
0112680201 & 2002-07-14 & 9924.8 & 9930.4 & 8787.2 & 3.83 $\pm$ 0.02 & 3.74 $\pm$ 0.02 & 21.67 $\pm$ 0.05 & Thin1 \\
0112680501 & 2002-07-25 & 1731.3 & 1730.1 & 1497.2 & 7.91 $\pm$ 0.07 & 7.94 $\pm$ 0.07 & 38.48 $\pm$ 0.16 & Thin1 \\
0125910501 & 2002-07-08 & 3387.5 & 3392.3 & 3099.9 & 2.04 $\pm$ 0.03 & 1.90 $\pm$ 0.03 & 14.61 $\pm$ 0.07 & Thin1 \\
0135940201 & 2002-07-10 & 1899.5 & 1903.4 & 1386.4 & 14.55 $\pm$ 0.09 & 14.75 $\pm$ 0.09 & 66.28 $\pm$ 0.22 & Thin1 \\
\hline     
\hline  
\end{tabular}
\tablefoot{The table includes, for each instrument, exposures and total rates in the matrix working ranges after background subtraction, together with the filter applied during the observation.\\
\tablefoottext{*}{PN data are relative to the medium filter }
}
\end{table*}

It is important to note that our matrices are a first example of response matrices for soft protons and have never been applied to observational data. This paper, then, represents the first example of spectral analysis of flares detected by {\it XMM-Newton}. Here, presenting our work and results in fitting each spectrum independently, we discuss  with particular attention the intercalibration between the MOS and PN matrices under the consideration that the input spectra measured by the three detectors must be statistically equivalent. 
The data reduction and a model-independent analysis of the considered observations are presented in Sect.~\ref{sect:observations} and  Sect.~\ref{sect:phenomenology}, respectively. The spectral analysis is described in Sect.~\ref{sect:spectral_analysis}; the results and systematics are shown in Sect.~\ref{sect:systematic} and discussed in Sect.~\ref{sect:discussion}. Conclusions are drawn in Sect.~\ref{sect:conclusion}.

\begin{table*} [t]
\caption{\label{tab:log_obs2} Log of the observations considered for matrix validation during the solar minimum activity.}                  
\begin{tabular}{c | c | c c c | c c c | c }     
\hline\hline       
 Observation ID&  \multicolumn{1}{c|} {Date}& \multicolumn{3}{c|} {Exposure} & \multicolumn{3}{c|} {2-11.5 keV Rate} & Filter \\
   & (y-m-d) &\multicolumn{3}{c|} {(s)} & \multicolumn{3}{c|} {(c/s)} & \\
   &   & MOS1 & MOS2 & PN  & MOS1 & MOS2 & PN  & \\
\hline 
0827241201 & 2019-12-13 & 9935.8 & 9949.2 & 8644.8 & 2.29 $\pm$ 0.02 & 1.45 $\pm$ 0.02 & 31.85 $\pm$ 0.06 & Thin1 \\
0840440201 & 2019-12-15 & 16784.5 & 16828.1 & 14585.8 & 2.21 $\pm$ 0.01 & 1.47 $\pm$ 0.01 & 28.32 $\pm$ 0.05 & Thin1 \\
0841680901 & 2019-12-06 & 13140.0 & 13148.3 & 10596.3 & 4.19 $\pm$ 0.02 & 2.44 $\pm$ 0.02 & 62.65 $\pm$ 0.08 & Medium \\
0842080601 & 2019-12-01 & 28707.5 & 28746.8 & 22545.8 & 6.98 $\pm$ 0.02 & 6.88 $\pm$ 0.02 & 61.32 $\pm$ 0.06 & Thin1 \\
0844210101 & 2019-12-21 & 17966.1 & 18019.9 & 14406.4 & 2.03 $\pm$ 0.01 & 1.37 $\pm$ 0.01 & 35.29 $\pm$ 0.05 & Medium\tablefootmark{*}\\
0844860101 & 2019-12-03 & 17729.9 & 17756.8 & 14797.5 & 3.64 $\pm$ 0.02 & 2.46 $\pm$ 0.01 & 42.75 $\pm$ 0.06 & Medium \\
0844860201 & 2019-12-09 & 8510.3 & 8776.8 & 4674.1 & 15.41 $\pm$ 0.04 & 11.77 $\pm$ 0.04 & 131.16 $\pm$ 0.17 & Medium \\
0844860301 & 2019-12-31 & 13241.4 & 13267.7 & 11618.7 & 2.19 $\pm$ 0.01 & 1.35 $\pm$ 0.01 & 29.25 $\pm$ 0.05 & Medium \\
0844860401 & 2020-01-02 & 6947.6 & 7180.7 & 3136.9 & 21.28 $\pm$ 0.06 & 15.23 $\pm$ 0.05 & 195.36 $\pm$ 0.25 & Medium \\
0844860601 & 2020-01-06 & 36583.7 & 36650.0 & 29393.4 & 4.88 $\pm$ 0.01 & 4.63 $\pm$ 0.01 & 42.38 $\pm$ 0.04 & Medium \\
0844860801 & 2020-01-12 & 7208.8 & 7216.6 & 6317.3 & 1.26 $\pm$ 0.02 & 0.79 $\pm$ 0.01 & 17.20 $\pm$ 0.06 & Medium \\
0845280401 & 2019-12-14 & 12914.8 & 12931.9 & 9846.0 & 5.90 $\pm$ 0.03 & 4.21 $\pm$ 0.03 & 67.68 $\pm$ 0.10 & Thin1 \\
0852190101 & 2019-12-30 & 6733.8 & 6808.4 & 3452.5 & 14.67 $\pm$ 0.05 & 11.33 $\pm$ 0.04 & 164.55 $\pm$ 0.22 & Medium\tablefootmark{*}\\
0854590401 & 2019-12-27 & 15987.3 & 15997.3 & 13638.2 & 2.61 $\pm$ 0.01 & 1.78 $\pm$ 0.01 & 28.93 $\pm$ 0.05 & Thin1 \\
   &   &      &      &     &      &      &     & \\
0827060801 & 2020-07-29 & 16698.6 & 16726.4 & 13327.3 & 5.20 $\pm$ 0.02 & 2.94 $\pm$ 0.02 & 38.81 $\pm$ 0.06 & Thin1\tablefootmark{**}\\
0862400101 & 2020-07-01 & 18276.5 & 18305.6 & 12727.6 & 9.40 $\pm$ 0.02 & 7.83 $\pm$ 0.02 & 78.60 $\pm$ 0.08 & Medium \\
0862730601 & 2020-07-13 & 20822.8 & 20865.3 & 14480.9 & 10.53 $\pm$ 0.02 & 8.74 $\pm$ 0.02 & 77.39 $\pm$ 0.07 & Medium \\
0863810301 & 2020-07-05 & 5699.6 & 5712.3 & 4652.3 & 4.58 $\pm$ 0.03 & 3.97 $\pm$ 0.03 & 43.29 $\pm$ 0.10 & Thin1 \\
0863880401 & 2020-07-20 & 2439.4 & 2443.5 & 2016.2 & 5.05 $\pm$ 0.05 & 4.13 $\pm$ 0.04 & 46.39 $\pm$ 0.16 & Medium \\
0865040201 & 2020-08-18 & 8882.3 & 8906.5 & 7788.4 & 2.01 $\pm$ 0.02 & 1.58 $\pm$ 0.02 & 24.39 $\pm$ 0.06 & Medium \\
0865040401 & 2020-08-28 & 6630.4 & 6651.5 & 5736.3 & 3.04 $\pm$ 0.02 & 2.55 $\pm$ 0.02 & 32.92 $\pm$ 0.08 & Medium \\
\hline                    
\hline  
\end{tabular}
\tablefoot{The table includes, for each instrument, exposures and total rates in the matrix working ranges after background subtraction, together with the filter applied during the observation.\\
\tablefoottext{*}{PN data are relative to the thin filter} \\
\tablefoottext{**}{PN data are relative to the medium filter}
}
\end{table*}

\section{Observations and data reduction}
\label{sect:observations}
Our analysis considered as the signal the events collected in the intervals with flaring activities and extracted the background from quiescent emission intervals.
{\it XMM-Newton} data were obtained from the archive,\footnote{\url{https://www.cosmos.esa.int/web/xmm-newton/xsa}} selecting observations with no strong sources in the field of view and full frame mode available either for MOS and PN, where the light curves and the focal plane images showed an evident presence of flares.  
These criteria were applied for data selection in four epochs: December 2001--January 2002 (named epoch $A$ and identified by the red color throughout the paper), July--August 2002 (epoch $B$, blue color), December 2019--January 2020 (epoch $C$, green color), and July--August 2020 (epoch $D$, gray color). They involve different environmental conditions due to seasons or to solar cycle phases, the years 2001--2002 being at the maximum of solar cycle 23 and 2019--2020 at the minimum of solar cycle 24. This choice would allow us to investigate any eventual effect on flares, as the dependence of its density on the angle between the satellite orbit and the magnetosphere axis  \citep{Fioretti2016}.
Table~\ref{tab:log_obs1} and Table~\ref{tab:log_obs2} present the log of the selected observations at the maximum and minimum solar activity, respectively. 

All data were reprocessed using the standard tools, {\it epproc} and {\it emproc} (SAS ver. 21.0.0). Light curves and spectra (both from flares and quiescent emission) were accumulated by selecting only single events (PATTERN=0) to be compliant with the available proton matrices. Bad pixels (XMMEA\_EM and XMMEA\_EP  set to on and FLAG=0) were excluded and no spatial selection was applied, again in agreement
with the matrices that are relative to the whole field of view. 

Flare spectra were accumulated, filtering EPIC events for periods of high background flaring activity,\footnote{\url{https://www.cosmos.esa.int/web/xmm-newton/sas-thread-epic-filterbackground}} while intervals with steady emission constitute our background. This includes the sources in the field of view, the diffuse X-ray sky, and the instrumental dark components. In particular, in the 50 s binned light curves of each observation, we first identified by eye periods with low steady rates (see the interval between the two vertical dashed lines in  Fig.~\ref{fig:src_selection} as an example), where we computed the average rate ($ave$) and the root mean square ($rms$) relative to it. These were used to produce good time intervals (GTIs) to accumulate spectra. Background GTIs were created by selecting intervals with rates lower than two times the  $rms$ above the average [$ave+2 \times rms$], where up to $\sim$95\% of the background fluctuations are included.  Flares GTIs instead are relative to periods where background fluctuations have a probability lower than $\sim$0.3\% of contaminating the signal [rate $>$ $ave+ 3 \times rms$].
The fraction of events above or below the two thresholds in units of $rms$ assumes a Gaussian distribution of the count rates.
Figure~\ref{fig:src_selection} shows the background interval for observation ID 0844860601 with two dashed vertical lines, and the rate thresholds for the background and for the flares with dash-dotted and dotted lines, respectively.

The average spectrum per observation was accumulated without taking into account eventual spectral variations in the observations. The spectral analysis was performed on  background-subtracted spectra with channels grouped in order to have a minimum of 50 counts per bin, allowing for the application of the $\chi^2$ statistic.

\section{Model-independent analysis}
\label{sect:phenomenology}

A model-independent analysis on the selected observational data is required before applying the response matrices, considering that our sample spans 20 years of satellite life. This was performed using the 2-11.5 keV flare rates measured by the three instruments in the four epochs after subtracting the contemporary background (see Table~\ref{tab:log_obs1} and Table~\ref{tab:log_obs2}).

It is important to note that data relative to MOS1 CCD3 and CCD6 are not available after 2005, due to damage from micrometeorite impacts. The main effects expected from  losing the two detectors should be an average reduction of MOS1 rates on the order of $\sim$30\%. However, looking at the rates listed in the two tables, a more complex behavior between different epochs appears.

We investigated, at first, the variation in the count rate between the three cameras in each epoch. To this purpose, the averages plus errors and the dispersion relative to them evaluated from the standard deviation ($rms$) were computed. From the resulting values,  presented in Table~\ref{tab:rates},  it is evident that there are no statistically significant variations in each instrument rate during epochs because of the large $rms$. 

However, the comparison of the average values shows some inconsistency between instruments:  PN average rates in the interval $C$ and $D$ are significantly higher with respect to those in $A$ and $B$, while the MOS2 ones are lower, even if no variation in this detector response is expected. In addition, the average rates observed by MOS1, where a decrease of $\sim$30\% is expected,  are almost constant, even if values in $A$ and $B$ are higher than those in $C$ and $D$. A more precise evaluation of these differences was computed from the ratios between rates in the three instruments. The corespondent average values,  listed in Table~\ref{tab:rate_ratio}, confirm the inconsistency.

\begin{table*} [th]
\caption{Mean and standard deviation ($rms$) in units of c/s for the rates of the analyzed flares for each instrument and each epoch.}             
\label{tab:rates}      
\centering          
\begin{tabular}{l c c c c c c c c }     
\hline\hline  
 & \multicolumn{2}{c}{$A$} & \multicolumn{2}{c}{$B$} & \multicolumn{2}{c}{$C$} & \multicolumn{2}{c}{$D$} \\
 & mean  & $rms$ & mean  & $rms$  & mean  & $rms$  & mean  & $rms$  \\
MOS1 & 7.3$\pm$1.7   & 6.4 & 8.3$\pm$1.3 & 5.8    & 6.4$\pm$1.6 & 6.0    & 5.7$\pm$1.1 & 2.9\\
MOS2  & 7.1$\pm$1.6  & 6.2 & 8.2$\pm$1.3 & 5.8    & 4.8$\pm$1.2 & 4.5    & 4.5$\pm$1.0 & 2.5\\
PN    & 33.2$\pm$7.0 & 27.2 & 36.6$\pm$4.6 & 20.2 & 67.0$\pm$14.3 & 53.7 & 48.8$\pm$7.4 & 19.6 \\
\hline\hline
\end{tabular}
\end{table*}

\begin{table} [th]
\small
\caption{Average of the rate ratios measured in each instrument for the four epochs. }             
\label{tab:rate_ratio}      
\centering          
\begin{tabular}{l c c c c }     
\hline\hline  
 & \multicolumn{1}{c}{$A$} & \multicolumn{1}{c}{$B$} & \multicolumn{1}{c}{$C$} & \multicolumn{1}{c}{$D$} \\
MOS1/MOS2 &  1.03$\pm$0.02 &  1.03$\pm$0.01 &  1.42$\pm$0.05 &  1.29$\pm$0.07\\
MOS1/PN   & 0.21$\pm$0.01  &  0.21$\pm$0.01 &  0.09$\pm$0.01 &  0.11$\pm$0.01\\
MOS2/PN   &  0.21$\pm$0.02 &  0.21$\pm$0.01 &  0.07$\pm$0.01 &  0.09$\pm$0.01\\
\hline\hline
\end{tabular}
\end{table} 

\begin{figure}
\centering
\includegraphics[width=5.5cm, angle= -90] {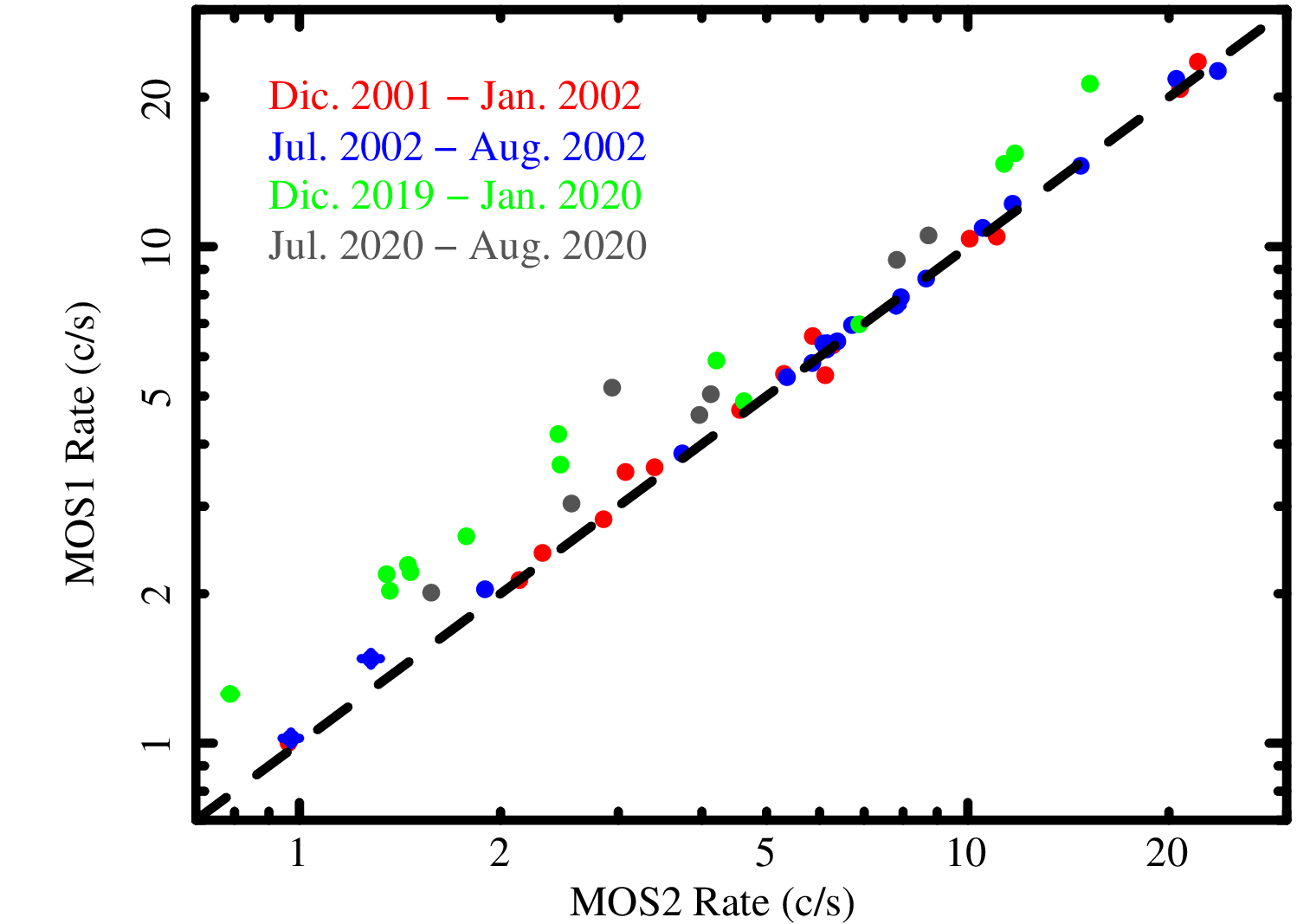}

\hspace{1.0cm}

\includegraphics[width=5.5cm, angle= -90]
{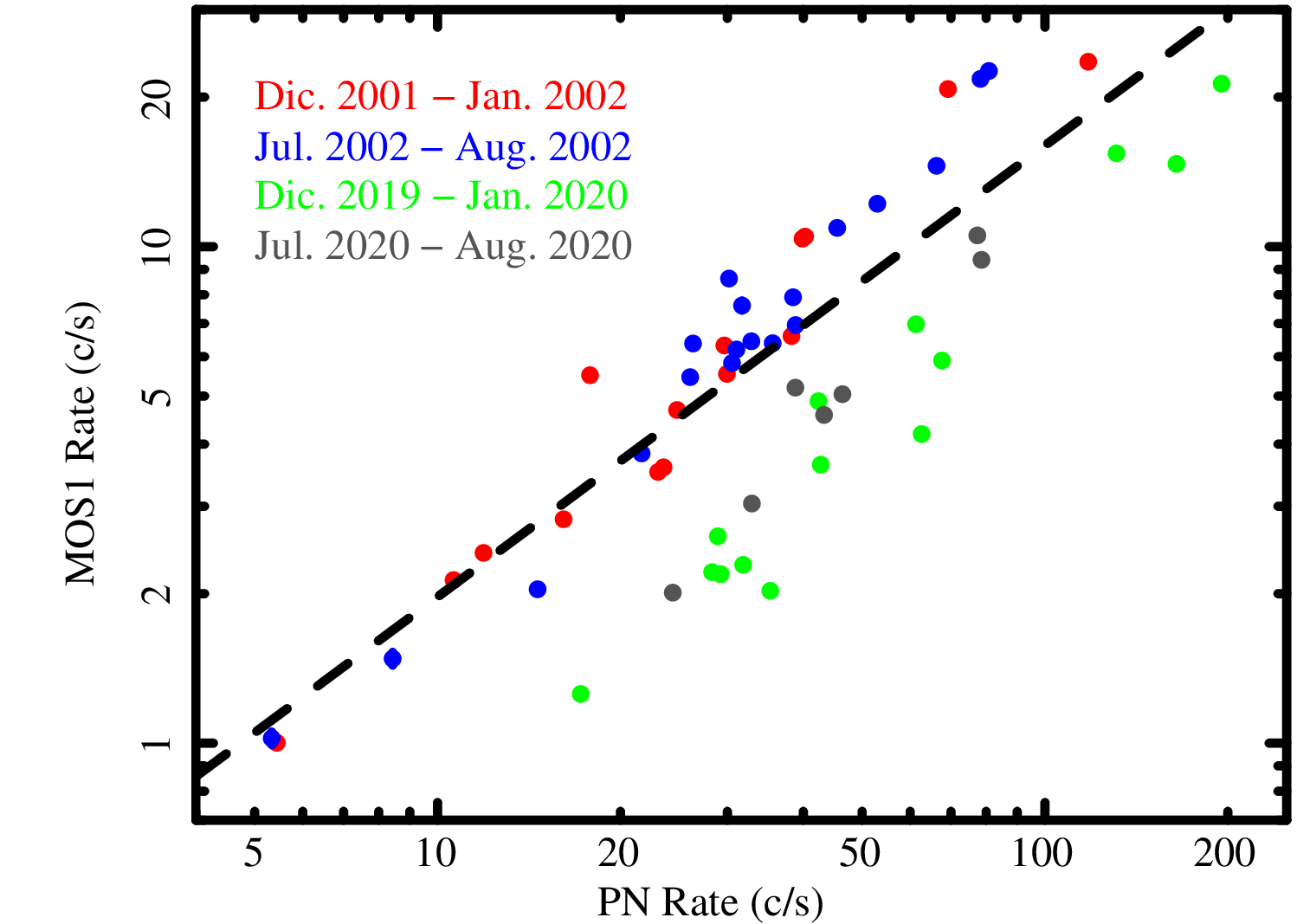}

\hspace{1.0cm}

\includegraphics[width=5.5cm, angle= -90]{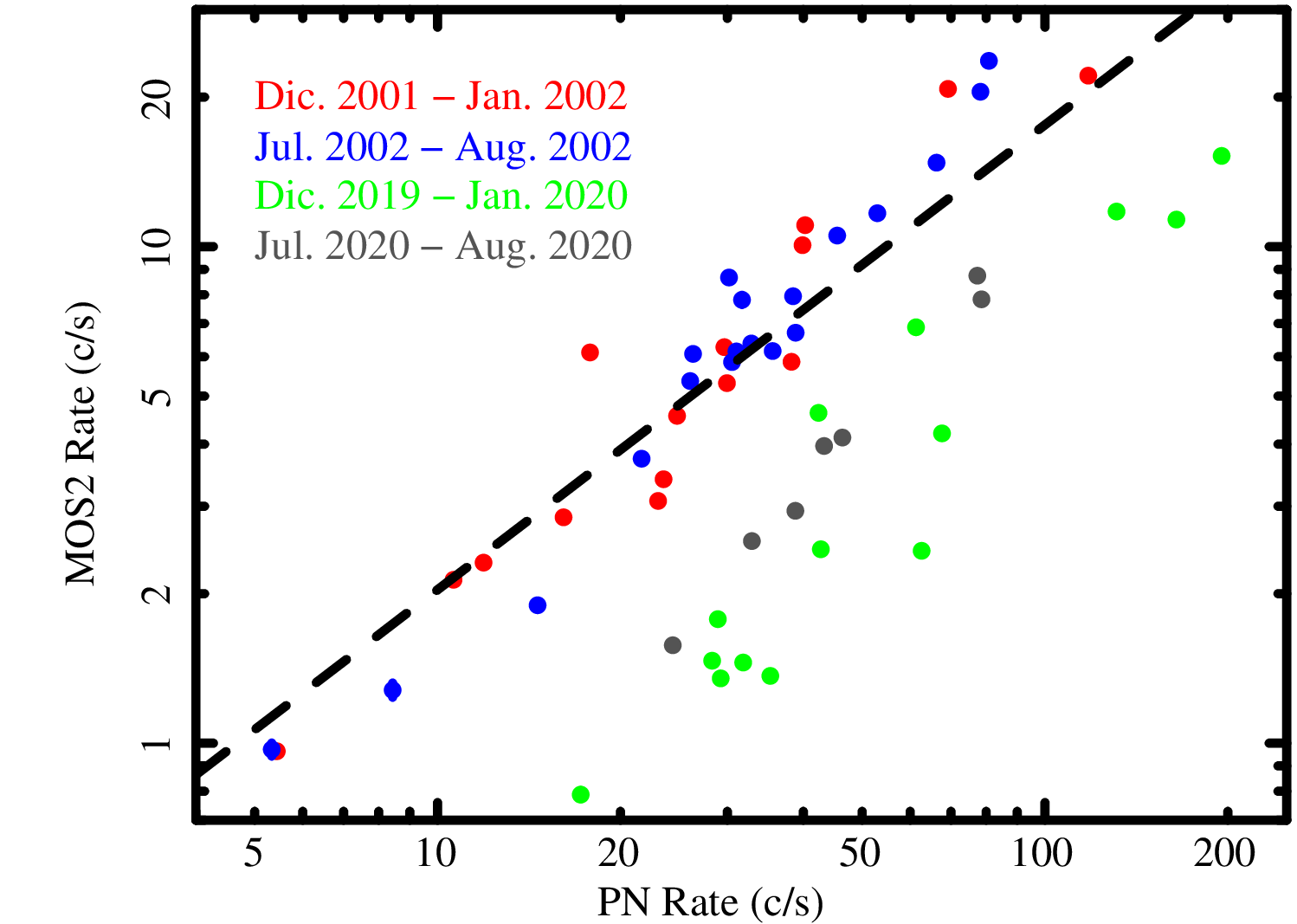}
\caption{Correlations between rates detected by the three instruments in the four epochs. Colors indicate epochs (see text).
The superimposed dashed black line is relative to the best fit of epoch $B$. The top panel is relative to MOS1 vs MOS2 rates, the middle panel to MOS1 vs PN, and the bottom panel to MOS2 vs PN. }
\label{fig:M2vsPN}
\end{figure}

As a further analysis, the correlations between the rates detected by the three instruments were investigated, plotting the values relative to MOS1 as a function of the corresponding MOS2 ones and the rates from MOS1 and MOS2 against the PN ones (see Fig.~\ref{fig:M2vsPN}).  
Each of the curves was then fit with a line fixing the intercept to zero ($y=k*x$). The fit used the least squares method for the minimization because the statistical errors are negligible with respect to the systematic ones.  The values of the resulting slopes, $k$, are shown in Table~\ref{tab:correlation_rate}; they are in agreement with the values in Table~\ref{tab:rate_ratio}, confirming again the inconsistencies detected between the rates in  epochs $C$ and $D$ and those in $A$ and $B$.

\begin{table*} [th]
\caption{Slope obtained by fitting with a line the MOS1 and MOS2 rates as function of MOS2 and PN rates (see the three panels of  Fig.~\ref{fig:M2vsPN}).}
\label{tab:correlation_rate}      
\centering          
\begin{tabular}{l c c c c }     
\hline\hline  
   & \multicolumn{1}{c}{$A$} & $B$ & $C$ & $D$ \\
\hline
\multicolumn{5}{c}{Proton Flares} \\
MOS1 vs MOS2 &  1.02$\pm$0.02  & 1.01$\pm$0.01 & 1.32$\pm$0.06 & 1.23$\pm$0.08\\
MOS1 vs PN   & 0.22$\pm$0.02  & 0.23$\pm$0.01 & 0.10$\pm$0.01 & 0.12$\pm$0.01\\
MOS2 vs PN   &  0.21$\pm$0.02  & 0.24$\pm$0.01 & 0.08$\pm$0.01 & 0.10$\pm$0.01\\
\\
\multicolumn{5}{c}{Photon Background} \\
MOS1 vs MOS2 &  0.99$\pm$0.01  & 0.98$\pm$0.01 & 0.677$\pm$0.005 & 0.71$\pm$0.02\\
MOS1 vs PN   & 0.36$\pm$0.03  & 0.38$\pm$0.02 & 0.248$\pm$0.003 & 0.25$\pm$0.01\\
MOS2 vs PN   &  0.36$\pm$0.03  & 0.39$\pm$0.01 & 0.368$\pm$0.005 & 0.36$\pm$0.01\\
\hline\hline
\end{tabular}
\end{table*}

As a check, correlations between rates detected by instruments were also investigated for photons using the background intervals identified in our analysis.   We computed these rates in the range of 2--5 keV to avoid contamination from the internal background component, which is dominant above this range\footnote{\url{https://xmm-tools.cosmos.esa.int/external/xmm\_user\_support/documentation/uhb/epicintbkgd.html}} and which is not affected by changes in the effective area. The results are shown in Table~\ref{tab:correlation_rate}. In this case, the correlation factors are  coherent with what was expected: the slope of the linear fit of MOS1 vs MOS2 rates in epochs $C$ and $D$ decreases by about 30\% with respect to the one in epochs $A$ and $B$, while no significant variation is observed for MOS2 vs PN rates.

\begin{figure}
\centering
\includegraphics[width=5.5cm, angle= -90] {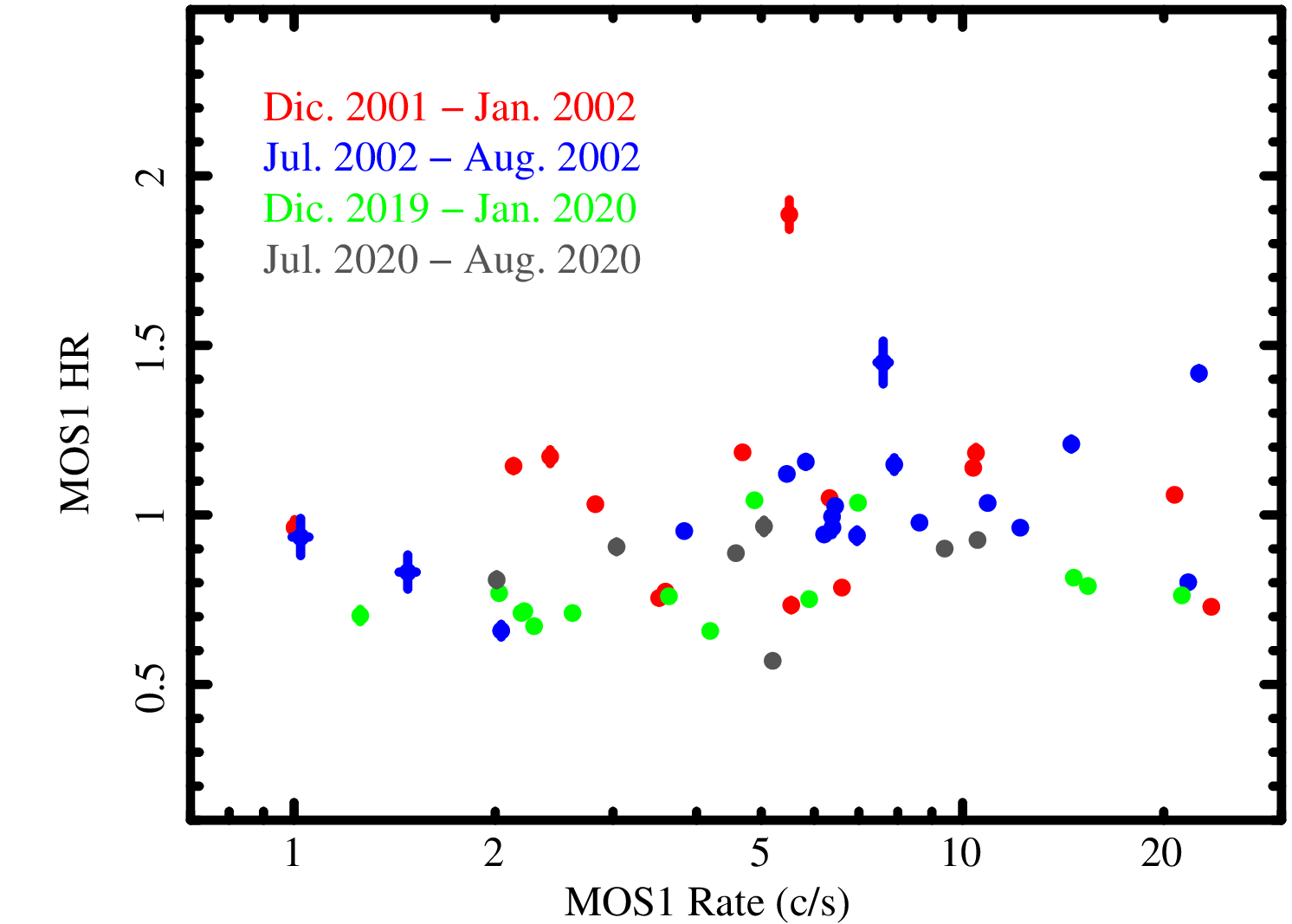}

\hspace{1.0cm}

\includegraphics[width=5.5cm, angle= -90]{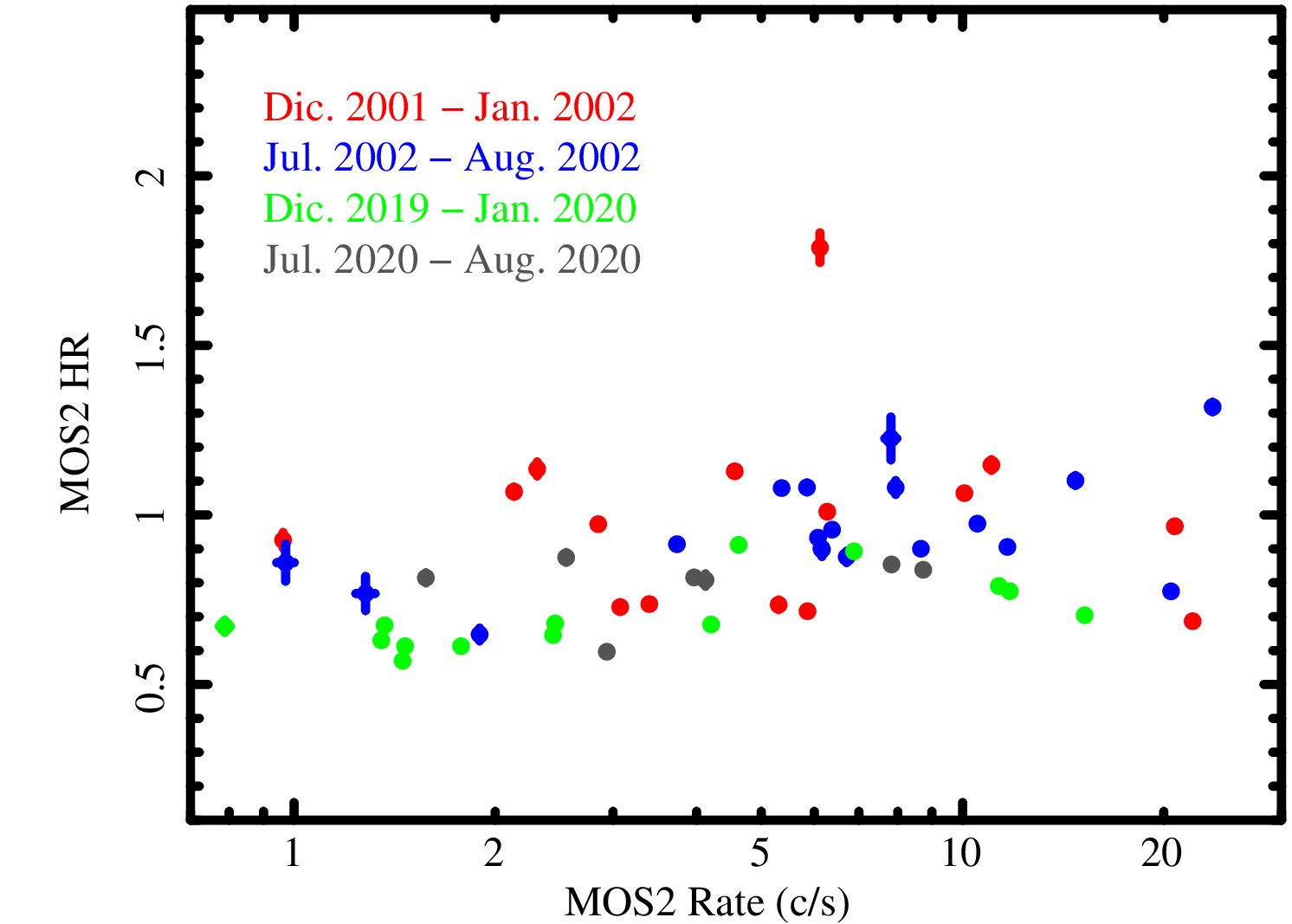}

\hspace{1.0cm}

\includegraphics[width=5.5cm, angle= -90]{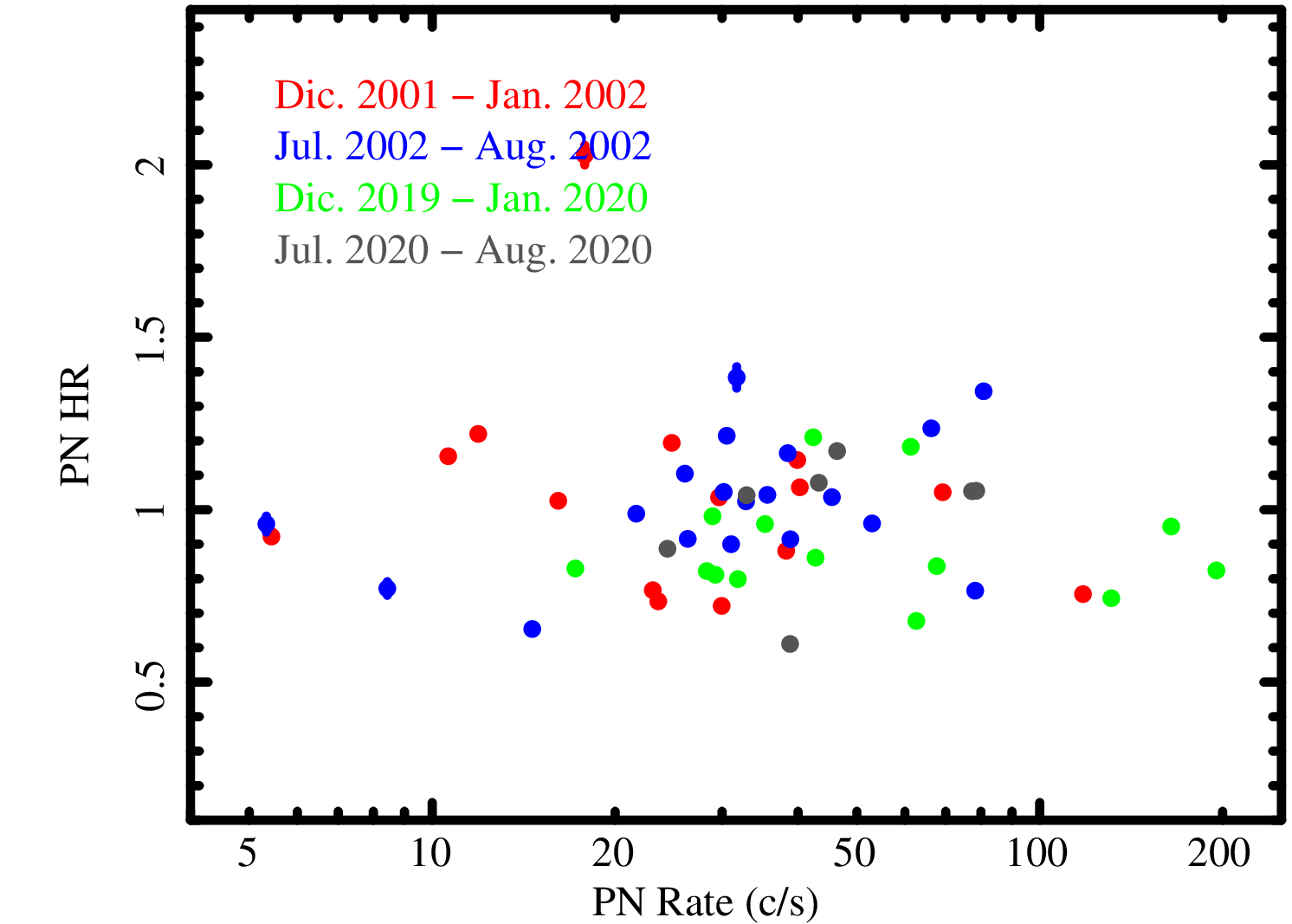}
\caption{Hardness ratio between counts in the ranges 5--11.5 keV and 2--5 keV vs. 2--11.5 keV rate for MOS1 (top panel), MOS2 (middle panel), and PN  (bottom panel). Colors indicate the four epochs (see text).}
\label{fig:hr_pn}
\end{figure}

To complete the model-independent analysis, an average hardness ratio was computed to investigate spectral differences among observations. The hardness was computed using counts in the ranges 5--11.5 keV and 2--5 keV.
Values as a function of the 2--11.5 keV rates for the three instruments
are shown in the three panels of Fig.\ref{fig:hr_pn}. Fitting all the points with a line, no correlations between  the hardness values and rates were obtained for both MOS1 and PN being the best fit value of the slope compatible with zero.
The MOS2 rate instead shows a weak positive correlation with hardness (slope=0.010$\pm$0.008).

\section{Flare spectral analysis}
\label{sect:spectral_analysis}

A large collection of {\it XMM-Newton} EPIC blank sky observations for the years 2000--2012 was provided by EXTraS, allowing one to characterize the focused background and to evaluate the fractions of time that the satellite spent in the regions of the Earth magnetosphere crossed by its orbit \citep{Ghizzardi2017}.
However, information on the input spectral model expected for the soft protons detected by {\it XMM-Newton} is still not accurate, despite the detailed studies performed on the subject \citep{Kronberg2020, Gastaldello2017, Kuntz2008}.

The ESA AREMBES project \citep{Macculi2021} predicted the soft proton environment for the Athena mission at the second Lagrange point. \citet{Lotti2018} and \citet{Fioretti2018} derived power law distributions above 50 keV for different regions of the magnetosphere and interplanetary space. 
These  models can also be considered valid for the {\it XMM-Newton} environment. 
\citetalias{Fioretti2024} presents the spectral analysis, using the proton response matrices, of EPIC MOS2 averaged spectra obtained from the EXTraS archive, normalized for the maximum intensity expected in 90\% of background observations \citep{Molendi2017}. The best-fitting model is in agreement with the predicted spectrum of the proton environment encountered by {\it XMM-Newton}'s orbit, predicted for the same 90\% exposure fraction. 

From our analysis, performed on single flares, we first started using power law models that describe well the proton environment above 50 keV.
However, because {\it XMM-Newton} EPIC detectors are also sensitive to lower-energy protons (see Fig.~\ref{fig:effarea}), alternative models must also be considered to investigate changes in the intrinsic slope. In this section, we show that a single power law is not able to fit the considered spectra and that, among all of the models tried, a power law plus a black body is the only one that better describes all of the spectra and gives coherent results between the three detectors.

\subsection{Fitting with a power law}
The analysis was performed within XSPEC (version: 12.12.1) and fitting all spectra individually in the common energy range of 2--11.5 keV. 
The $\chi^2$ statistic was applied to define the goodness of the fits. However, the reduced $\chi^2$ is not a good estimator because the selected spectra have different number of grouped channels (from 200 to 2000) depending on the statistics.
We instead used the null-hypothesis probability (NHP) provided by XSPEC, which is the probability of obtaining the observed value of $\chi^2$ if the model is a correct representation of the data.

We defined as acceptable fits those with NHPs higher than 2.7$\times$10$^{-3}$,  which corresponds to 3 $\sigma$ of a Gaussian distribution. Errors are reported at a 90\% confidence level.

The power law model is not a valid model for most of the spectra, given that 72\% of the 165 spectra have an NHP lower than the adopted threshold.  Among those consistent with a single power law, about 60\% are detected by the PN, and only in 6 observations of the 55 selected do the three detectors agree in observing a single power law. Their observation IDs are: 0037981201, 0037981601, 0037982701, 0111281501, and 0111281601 in interval $B$ and 0844860801 in interval $C$. These spectra are generally characterized by low count rates so that the larger errors allow us to fit with a single power law.

Reducing the fitting energy range to 5--11.5 keV significantly improved the fits.  In this case, 96\% of the spectra are  consistent with a single power law model, having an NHP within the 3 $\sigma$ Gaussian threshold.  The remaining 4\% spectra have in any case a probability within 4.5  $\sigma$.  
The measured spectral indices range in the intervals of 2--7 for MOS and 1--6 for PN and, on average, the values in epochs $A$ and $B$ are lower that those in $C$ and $D$. We observe no correlation between indices and rates. 

\begin{figure}
\centering
\includegraphics[width=5.5cm, angle= -90] {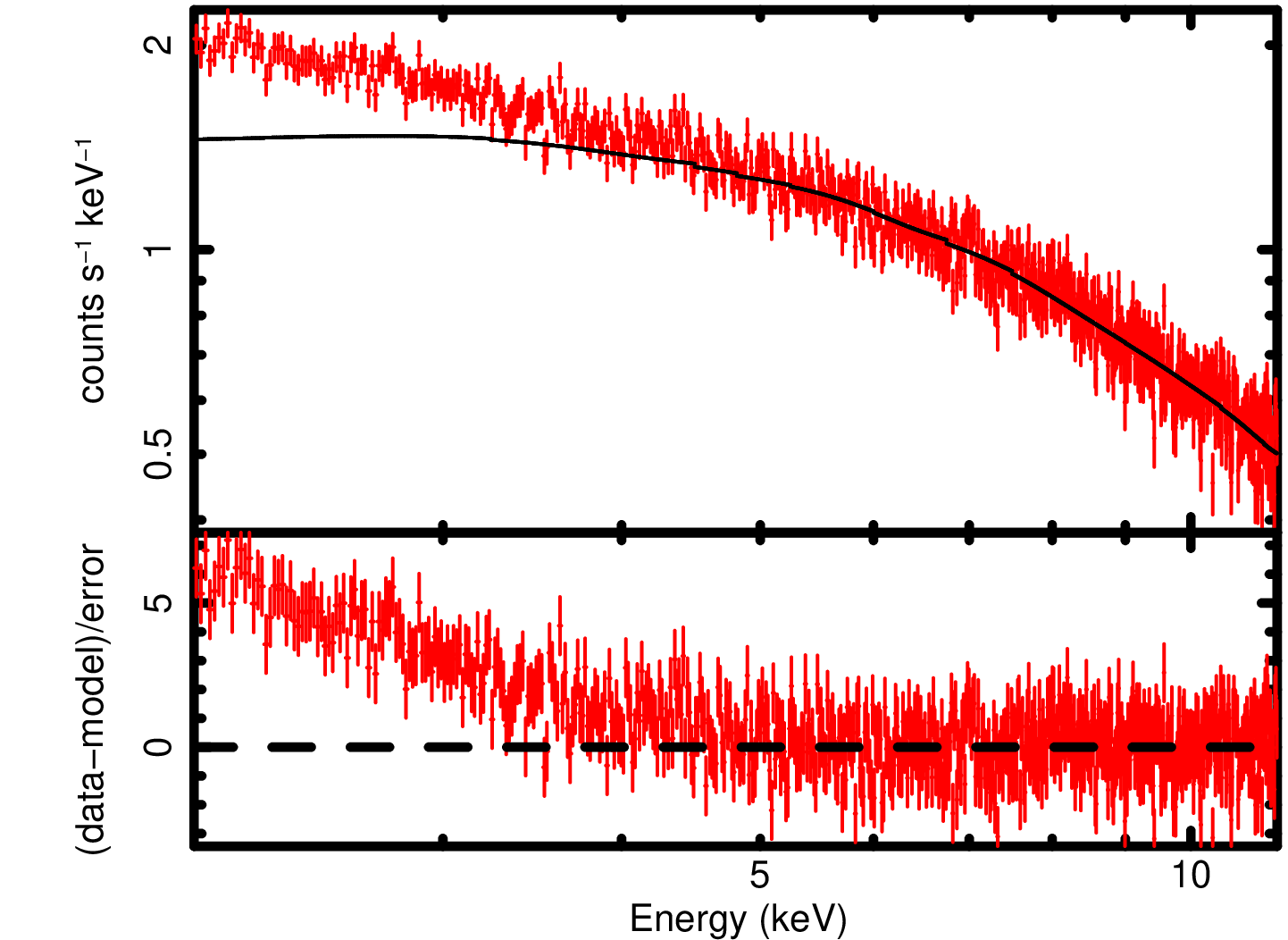}

\hspace{1.0cm}

\includegraphics[width=5.5cm, angle= -90] {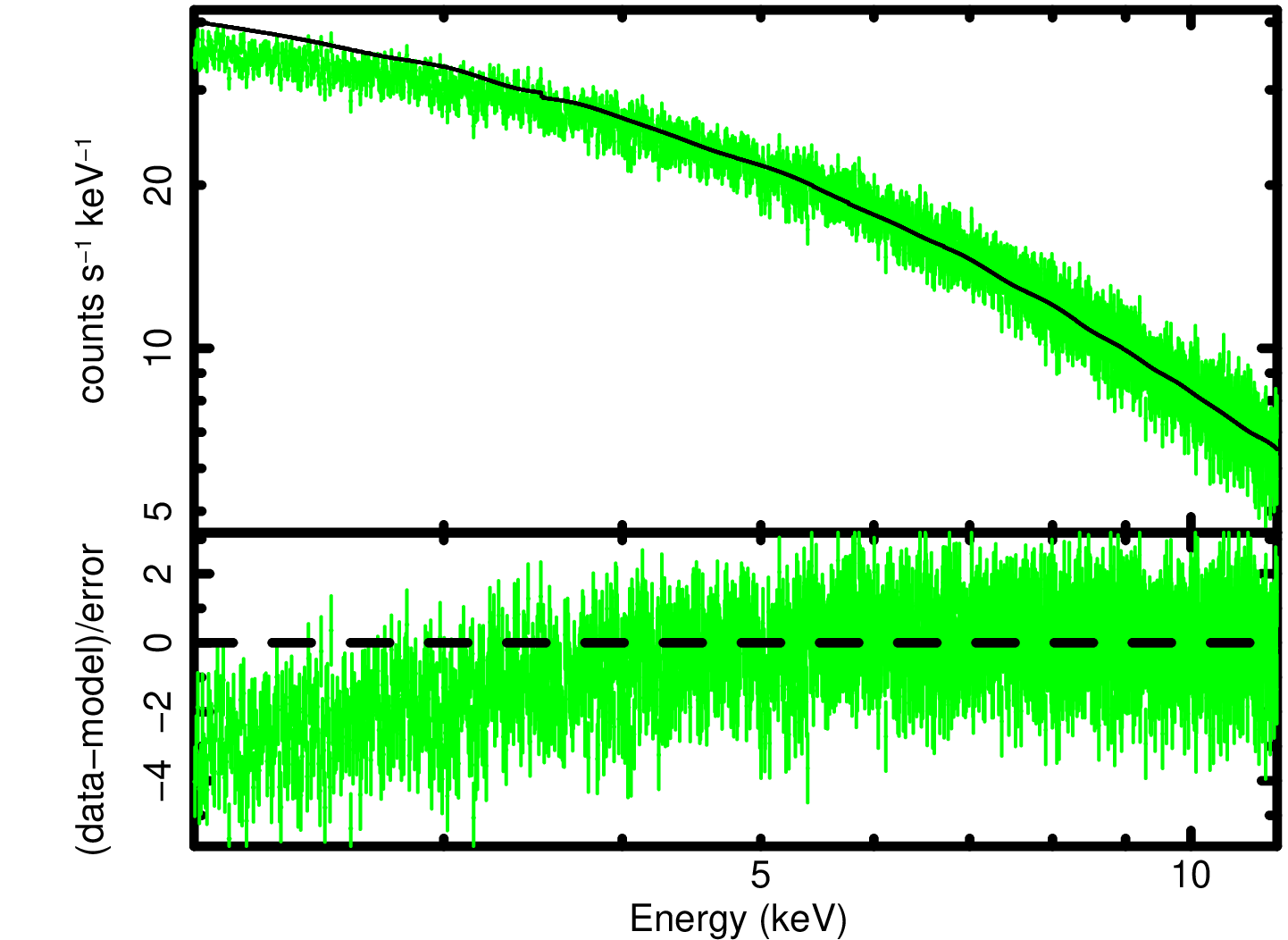}
\caption{Spectral fit with a power law.
The upper plot shows the  MOS1 spectrum relative to observation ID 0108060601. The top panel presents with a continuous line the 5--11 keV best-fit power law extrapolated down to 2 keV, while residuals over the whole range are presented in the lower panel. The lower plot is relative to the same figure for the 0852190101 PN spectrum.} 
\label{fig:spt_excess}
\end{figure}

In order to investigate possible alternative models for the whole energy range, we computed, for each spectrum and each instrument, the differences between the observed rates in the 2--5 keV band and the rates derived from the extrapolations of the 5--11.5 keV best-fitting power law. We found that for all MOS spectra and for most of the PN ones, the power law significantly underestimates the detected rates with statistically significant differences ($>$3 $\sigma$). As an example, Fig.~\ref{fig:spt_excess} shows in the upper plot the extrapolation of the 5--11.5 keV power law up to 2 keV for the MOS1 spectrum of observation ID 0108060601, where a low energy excess is evident.

For the PN spectra where the extrapolation does not underestimate the rates, some (0112551101 in the epoch $A$,  0037981601, 0037981701, and 0056022201 in $B$, 0844860301 and 0844860801 in $C$, and 0863810301 in $D$), present an excess consistent with zero and, in these cases, PN spectra are those fit by a single power law in the whole 2--11.5 keV range.   
In a few other observations (0052140201 and 0108061901 in epoch $A$,  0827241201,  0844210101,  0852190101, and 0854590401 in $C$, and 0862400101 in $D$),  the extrapolation of the power law led to higher rates than the detected ones. In the lower plot of Fig.~\ref{fig:spt_excess}, an example one of these cases is presented. For these spectra, a different spectral model should be considered, and consequently we excluded them from the following analysis. 

\begin{figure}
\centering
\includegraphics[width=5.5cm, angle= -90] {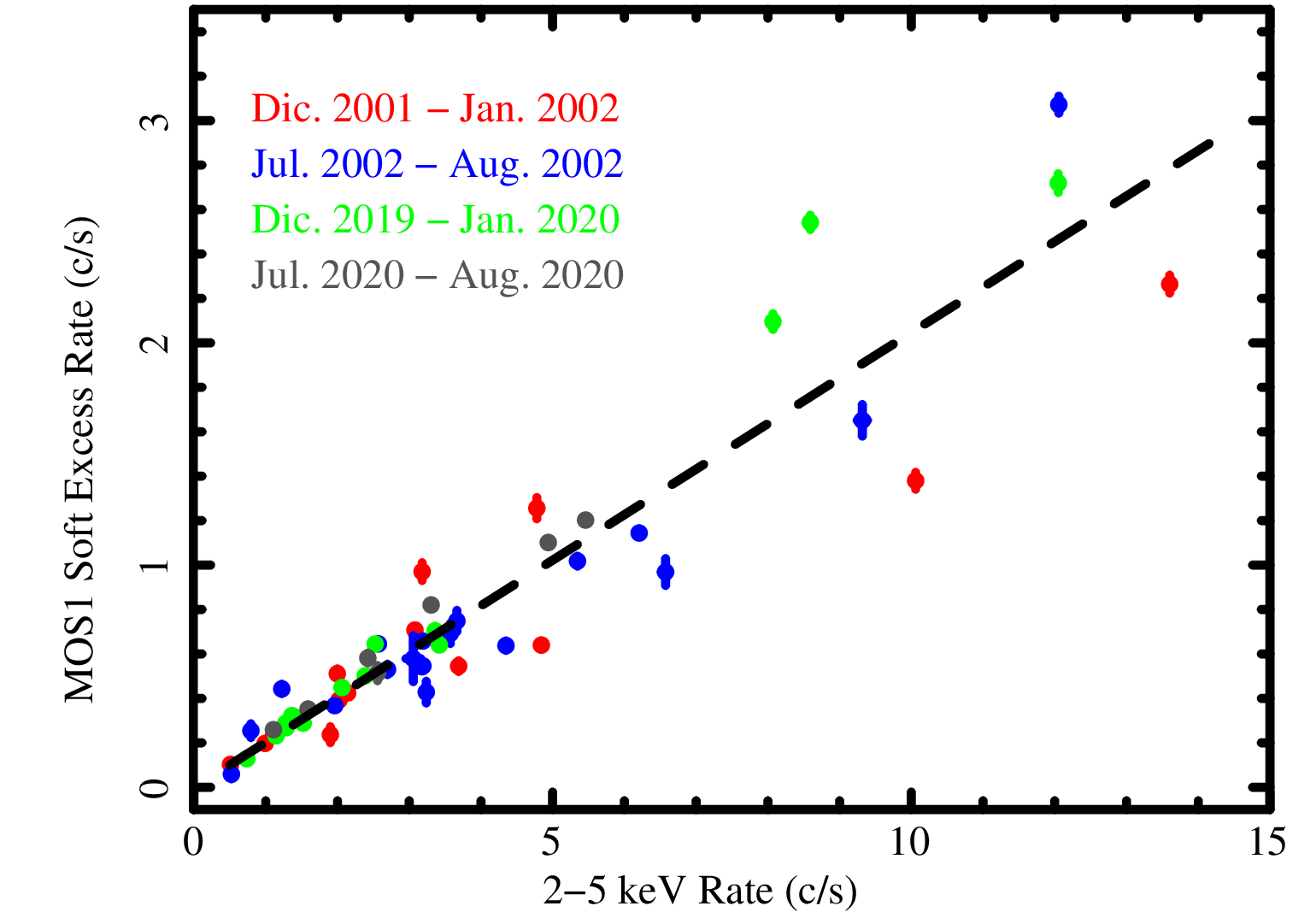}

\hspace{1.0cm}

\includegraphics[width=5.5cm, angle= -90]{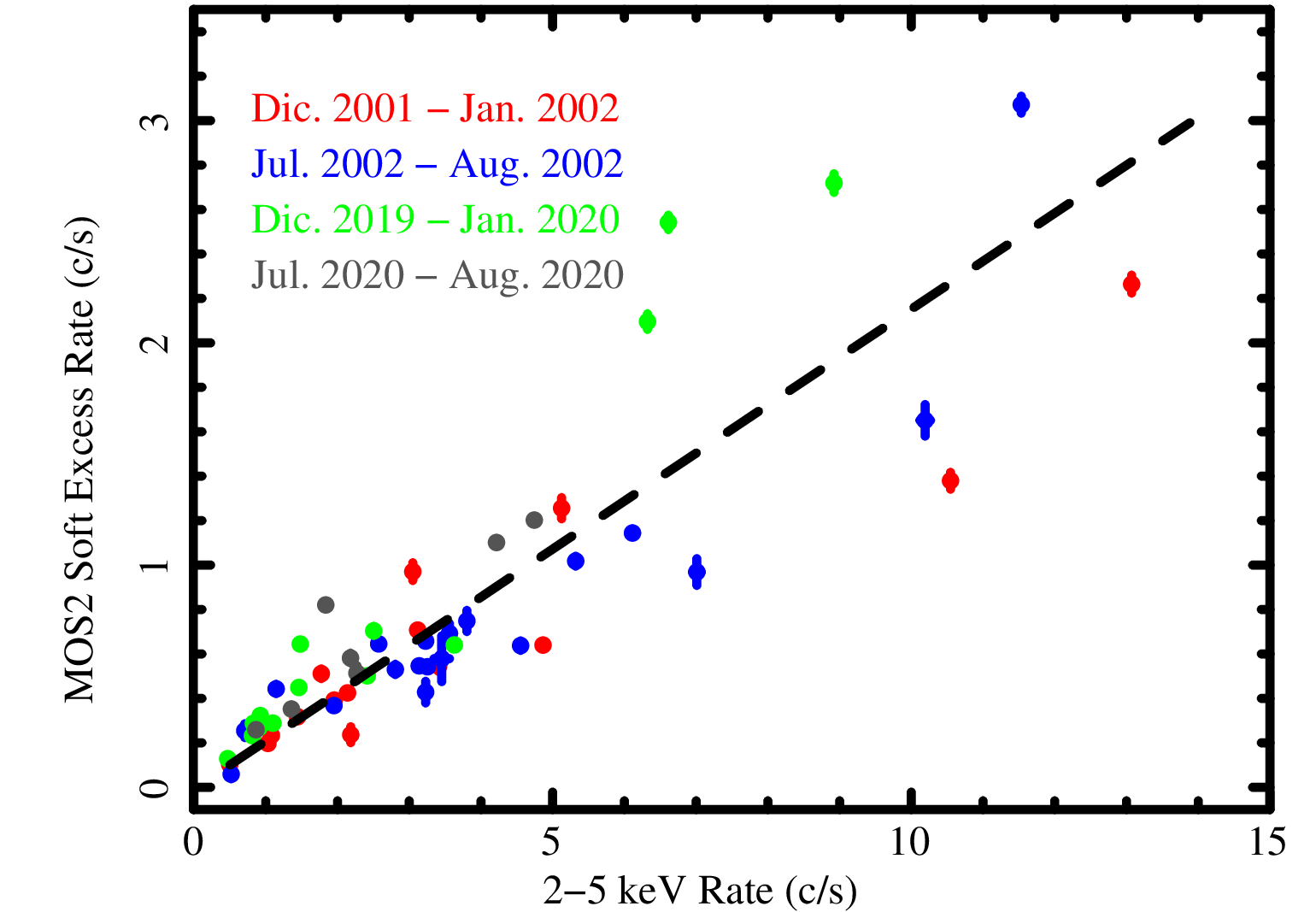}

\hspace{1.0cm}

\includegraphics[width=5.5cm, angle= -90]{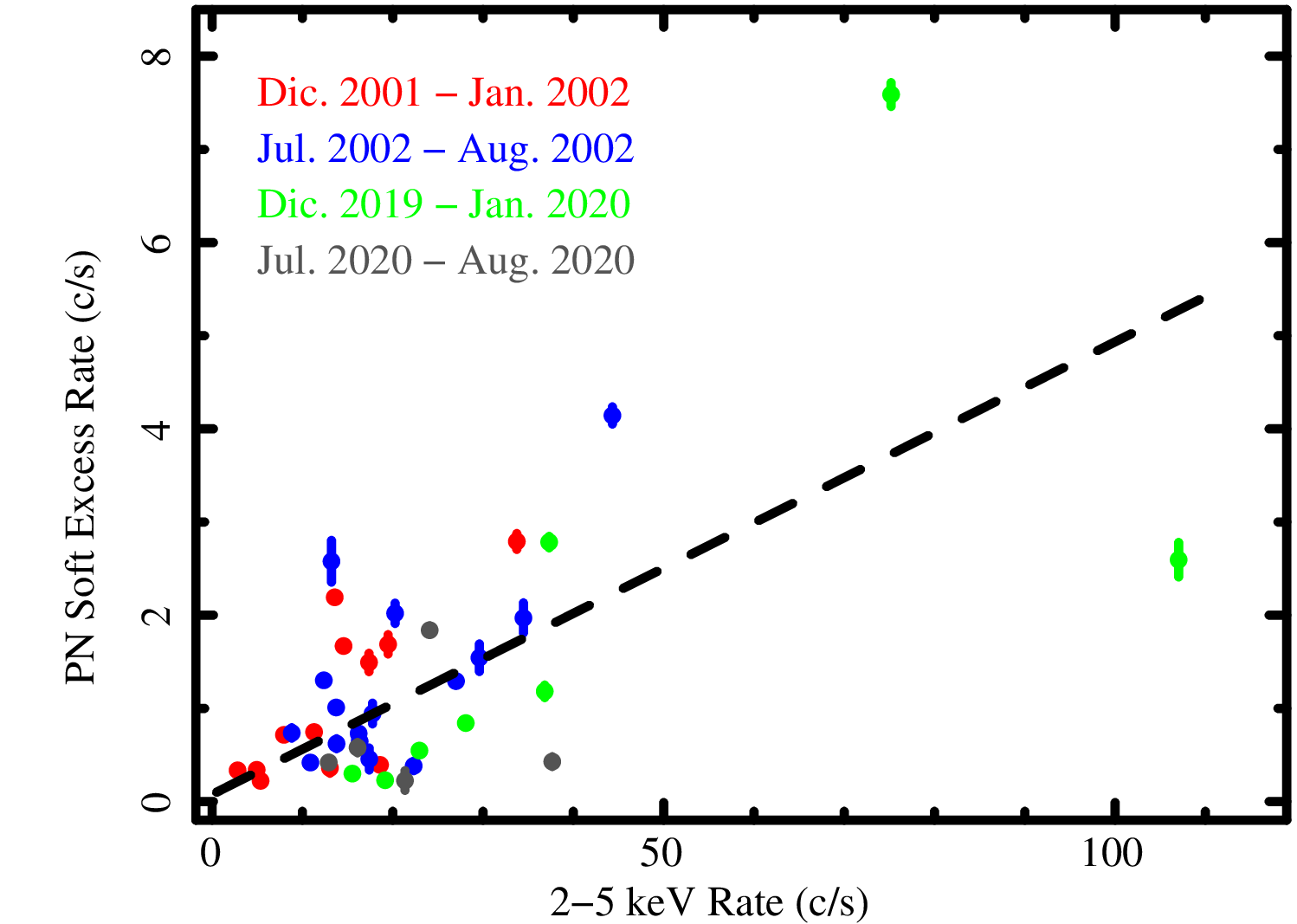}
\caption{Excess rates with respect of the 5--11.5 keV best-fit power law in the range of 2--5 keV as a function of the total 2--5 rates for MOS1 (top panel), MOS2 (middle panel), and PN (bottom panel). }
\label{fig:excess_pn}
\end{figure}

The excess rates were then correlated with the best-fit spectral indices and with the observed 2--5 keV rates.  While no significant correlation was observed with the spectral indices for any of the three detectors,
the correlation with the observed rates is instead well evident, with no significant differences between epochs.  
Figure~\ref{fig:excess_pn} shows the 2--5 keV excess rate with respect to the 5-11.5 keV power law plotted as functions of the 2--5 keV rates for the three instruments. MOS1 and MOS2 show a strong correlation, while the correlation is weaker for PN data. From fitting all points with a line, we found that the 2--5 keV excess rate is 21\% of the detected rate for MOS1 and MOS2, while it is only 5\% for the PN. 

\begin{figure}
\centering
\includegraphics[width=5.5cm, angle= -90] {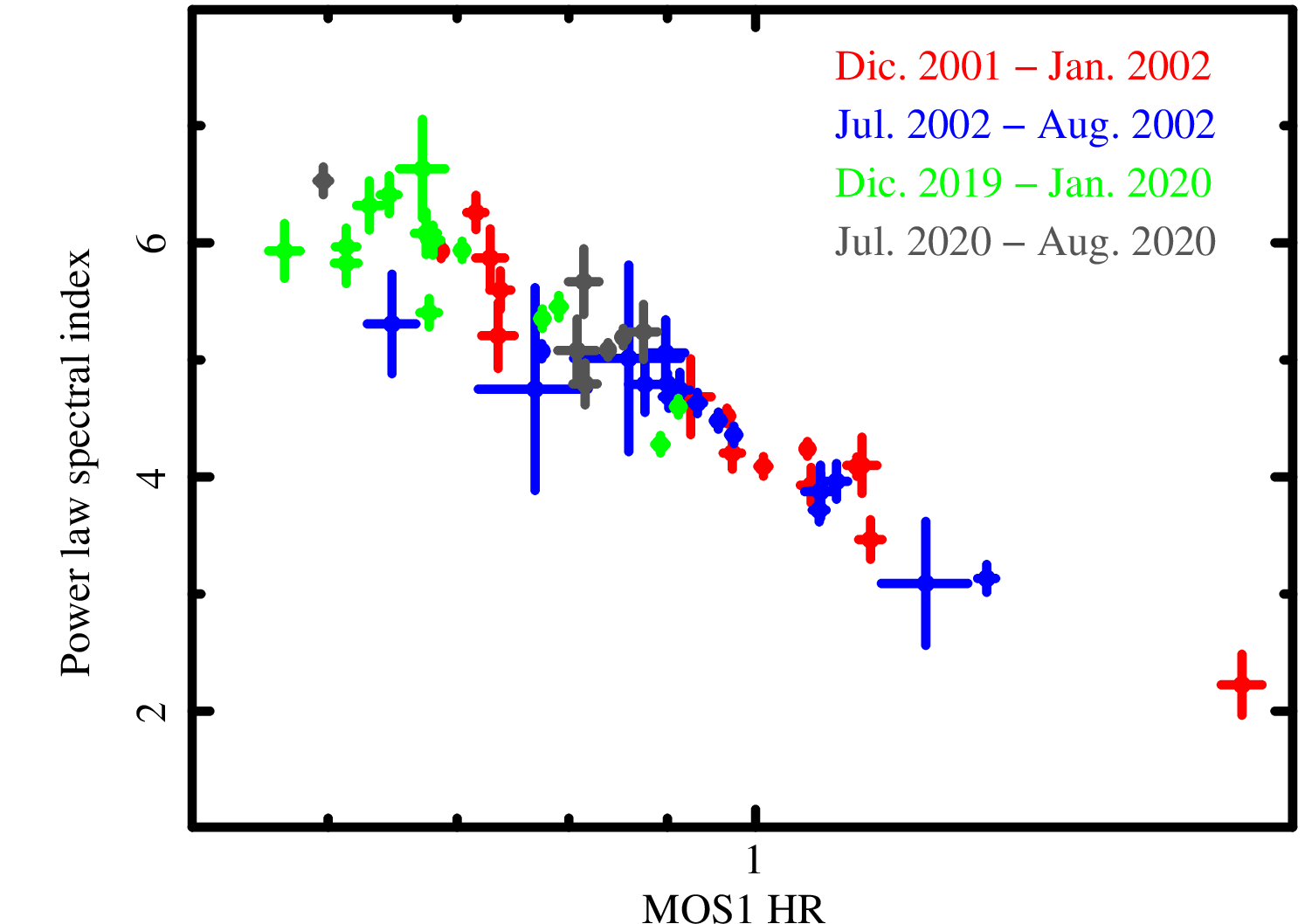}

\hspace{1.0cm}

\includegraphics[width=5.5cm, angle= -90]{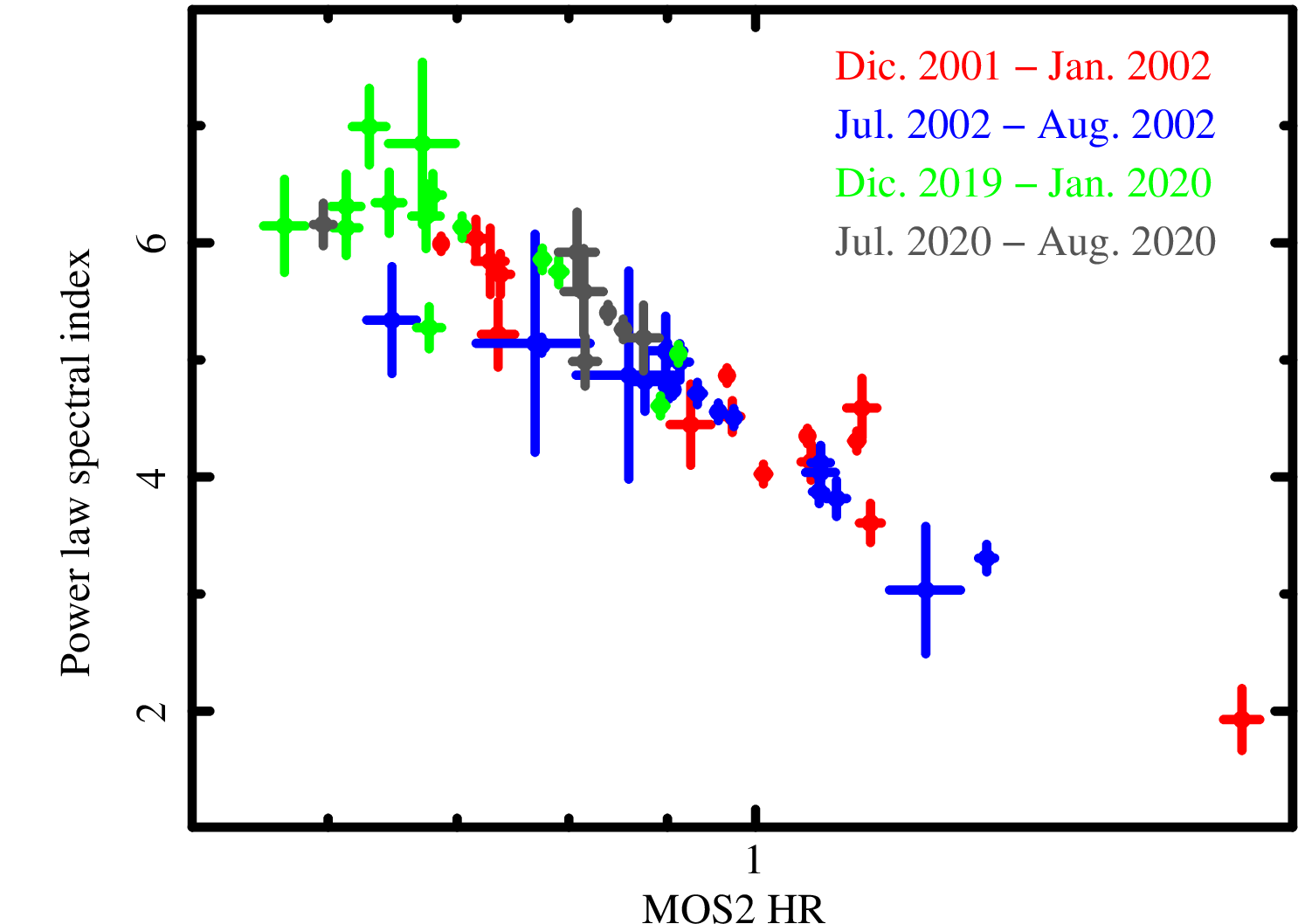}

\hspace{1.0cm}

\includegraphics[width=5.5cm, angle= -90]{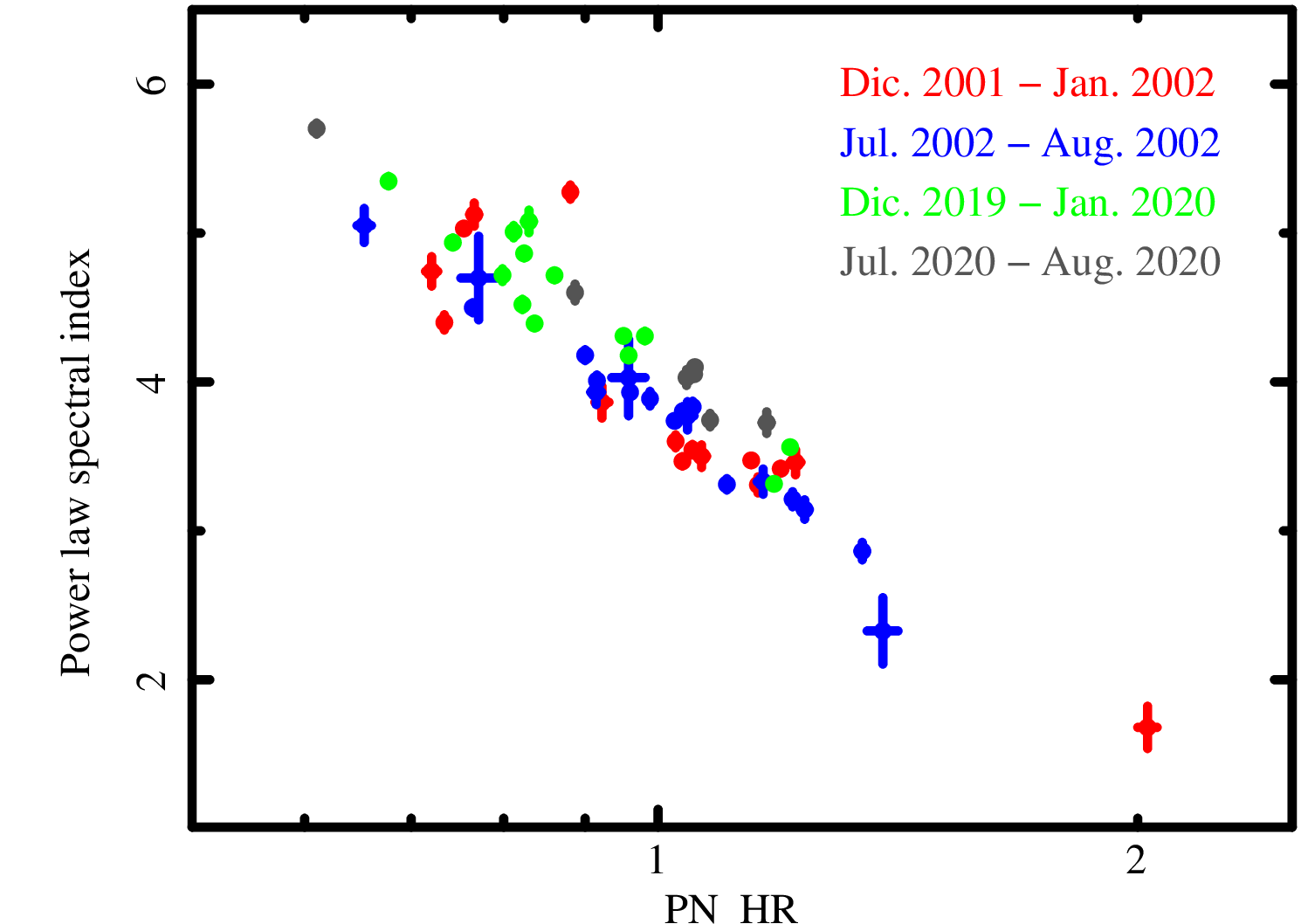}
\caption{Power law spectral index computed from fitting the 5-11.5 keV ranges vs. the hardness ratios for MOS1 (top panel), MOS2 (middle panel), and PN (bottom panel). }
\label{fig:alfahr_pn}
\end{figure}

The power law obtained from fitting in the 5--11.5 keV range extrapolated to lower energies constitutes the most of the detection for almost all spectra, the residuals comprising a small fraction of the total rate.
Taking into account this, and considering that this is a first check of the proton matrices with real data, to verify that the matrices correctly represent the spectral variations measured by the detectors, the best-fit values of the spectral indices are correlated with the hardness ratios. 
Figure~\ref{fig:alfahr_pn} shows the correlation plots for the three detectors.  It is evident that the spectral indices derived from the matrices closely correlate with the hardness ratios, with higher spectral indices corresponding to lower hardness ratios, and again no systematic differences are present between the four epochs.

\subsection{Fitting with a power law plus a second component}
According to the results presented in the previous subsection, a correct spectral modeling of the whole range must take into account the small excess at low energy.
The 2--11.5 keV spectra were then fit, adding a second component: a second power law and black body were investigated.
The fitting was performed in two steps. The statistics of the data, in fact, is not sufficient to limit the second component to the small fraction of the excess below 5 keV. 
Leaving all the parameters free in the 2--11.5 range, a strong reduction in the 5--11.5 keV power law contribution and a loss of correlation between the MOS and PN spectral indices were caused. The  power law was then fit, at first, limiting the range to 5--11.5 keV, and then the second component was added to the model after fixing the power law parameters. 

\subsubsection{Power law plus black body}
Using a black body as a second component,  83\% of the spectra have NHPs $>$ 2.7$\times$10$^{-3}$ and, in any case, the goodness of the fit improves for all of the spectra,  increasing the values of the NHP, even for those compatible with a single power law model. 
The MOS spectra not compatible with this model have  positive residuals below 3 keV, showing a more complex energy-dependent structure of the excess with respect to the simple black body model that we adopted. Most of these spectra have high count rates (0093160201 in epoch $B$, 0844860201 and 0844860401 in $C$, and 0827060801, 0862400101, and  0862730601 in $D$), but this effect is also present in the MOS2 spectrum 0844860601 with a moderate count rate.

For spectra with an acceptable fit, the $kT$ values are always around 1 keV. Specifically, the two MOSs present an average value of 1.0 keV with an $rms$ of 10\%, while for the PN a slightly lower average value (0.9 keV) is obtained, coupled with a larger standard deviation of the best-fitting values  (20\%). Fitting the best-fit values with a line, no correlation either with the excess rate or with the spectral indices is observed for this parameter in MOS data. A weak correlation with the excess rate is instead detected for the PN (slope=0.038$\pm$0.011). 

\subsubsection{Fitting with two power laws }
The composite model of two independent power laws is representative of only 37\% of the spectra. 

\subsection{Fitting with a power law with an energy-dependent spectral index}
In order to investigate the possibility that the low energy excess with respect to the power law is the effect of a spectral index steepening, we fit all spectra with a broken power law and with a power law whose index varies with energy as a log parabola \citep{Massaro2004}. Both  models give input proton spectra that cannot be considered physically reliable.

\subsubsection{Broken power law}
One of the simplest models with a variable spectral index is the broken power law that foresees two constant spectral indices, respectively, above and below the break energy, $E_{break}$. In our dataset, this model gives reduced $\chi^2$ lower than 2.7$\times$10$^{-3}$ for 86\% of the spectra.  The energy indices detected above $E_{break}$ are consistent within errors with those computed in the 5-11.5 keV energy range with a simple power law in all epochs except in $D$, where MOS detectors shows a discrepancy of about 10\%. However, all $E_{break}$ values are close to the lower boundaries of the soft proton matrices. Their best-fit values are in the range of 22--27 keV for MOS  and 10--19 keV for PN, with the consequence that the fitting procedure is not able to constrain the energy indices below $E_{break}$. 
Applying this model to the more statistically significant MOS2 spectra relative to 90\% of the observing time of the flaring background used to validate the matrices gives unacceptable $\chi^2$ (14855 for 1896 d.o.f) and underestimates the soft excess.

\subsubsection{Log-parabola power law}
The other model with an energy-dependent spectral index that we investigated was proposed by \citet{Massaro2004} for X-ray nonthermal emission. It follows the spectral distribution
\begin{equation}
F(E) = K \times E^{−(\alpha \, +\,  \beta \, \times \,  log (E) ) }
\label{eq:logpar}
,\end{equation}

\noindent
where the parameter $\alpha$ is the photon index at 1 keV and $\beta$
measures the curvature of the parabola.
The energy-dependent photon index, $\Gamma(E)$, formula, given by the log-derivative of Eq.~\ref{eq:logpar}, is
\begin{equation}
\Gamma(E) = \alpha + 2 \, \beta \,\times \, log (E)
\label{eq:index}
.\end{equation}

\begin{figure}
\centering
\includegraphics[width=5.5cm, angle= -90] {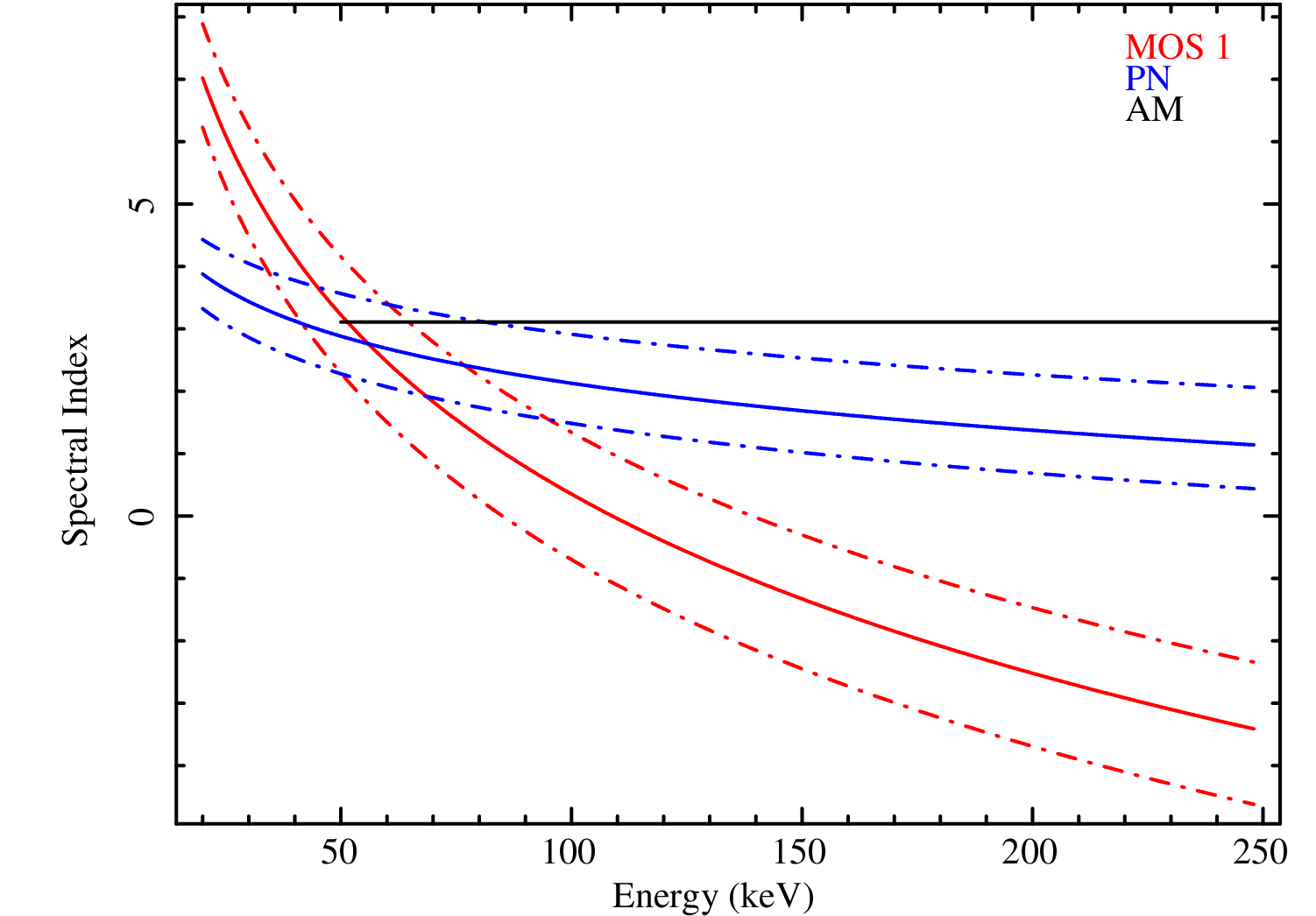}
\caption{Spectral index computed with Eq.~\ref{eq:index} for  MOS1 (continuous blue line) and  PN (continuous red line) spectra of observation ID 0110980601. The dash-dotted lines indicate the 90\% confidence level. For comparison, the spectral index relative to the 90\% flux in the active magnetosphere given in \citet{Lotti2018} is indicated  with the black line.    }
\label{fig:logpar}
\end{figure}

\noindent
This model fits 68\% of the 165 spectra well, but only for 51\% of the observation IDs is this true in all three detectors at the same time.
In these 51\% of cases, the parameters $\alpha$ and $\beta$ measured by the two MOSs are always consistent with  each other, while no coherence with the PN ones is obtained. Furthermore, computing the spectral indices over the matrix energy working range shows that above 50 keV the spectral indices cannot be considered in any way constant.  Figure~\ref{fig:logpar}
shows the results for the spectral index measured by MOS1 (blue line) and by PN (red line) for observation ID 0110980601 (epoch $D$) computed with Eq.~\ref{eq:index}. From the plot, it is evident that the two instruments measure completely different $\alpha$ and $\beta$ values and that the spectral index is not constant above 50 keV. 

\section{Spectral analysis systematics}
\label{sect:systematic}
To test the validity of the results and to quantify the systematic between the matrices, we compared the best-fit parameters obtained by the three detectors by analysing their ratios in pairs. This approach allows us to take into account errors detected by both detectors, while a simple fitting of one parameter with respect to the other would only account for errors on the fit variable.

\subsection{Power law}

\begin{figure}
\centering
\includegraphics[width=5.5cm, angle= -90] {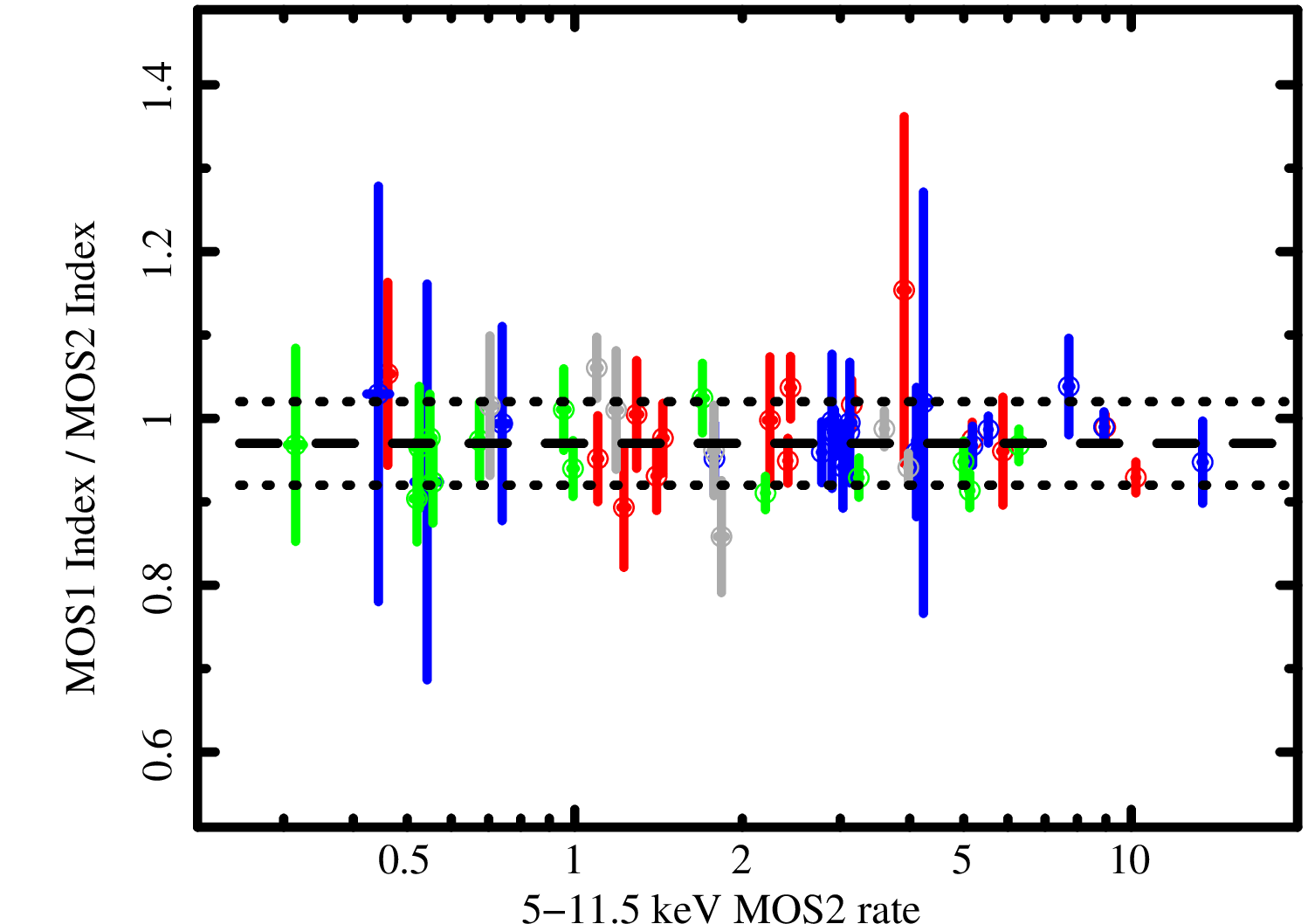}

\hspace{1.0cm}

\includegraphics[width = 5.5cm, angle= -90]{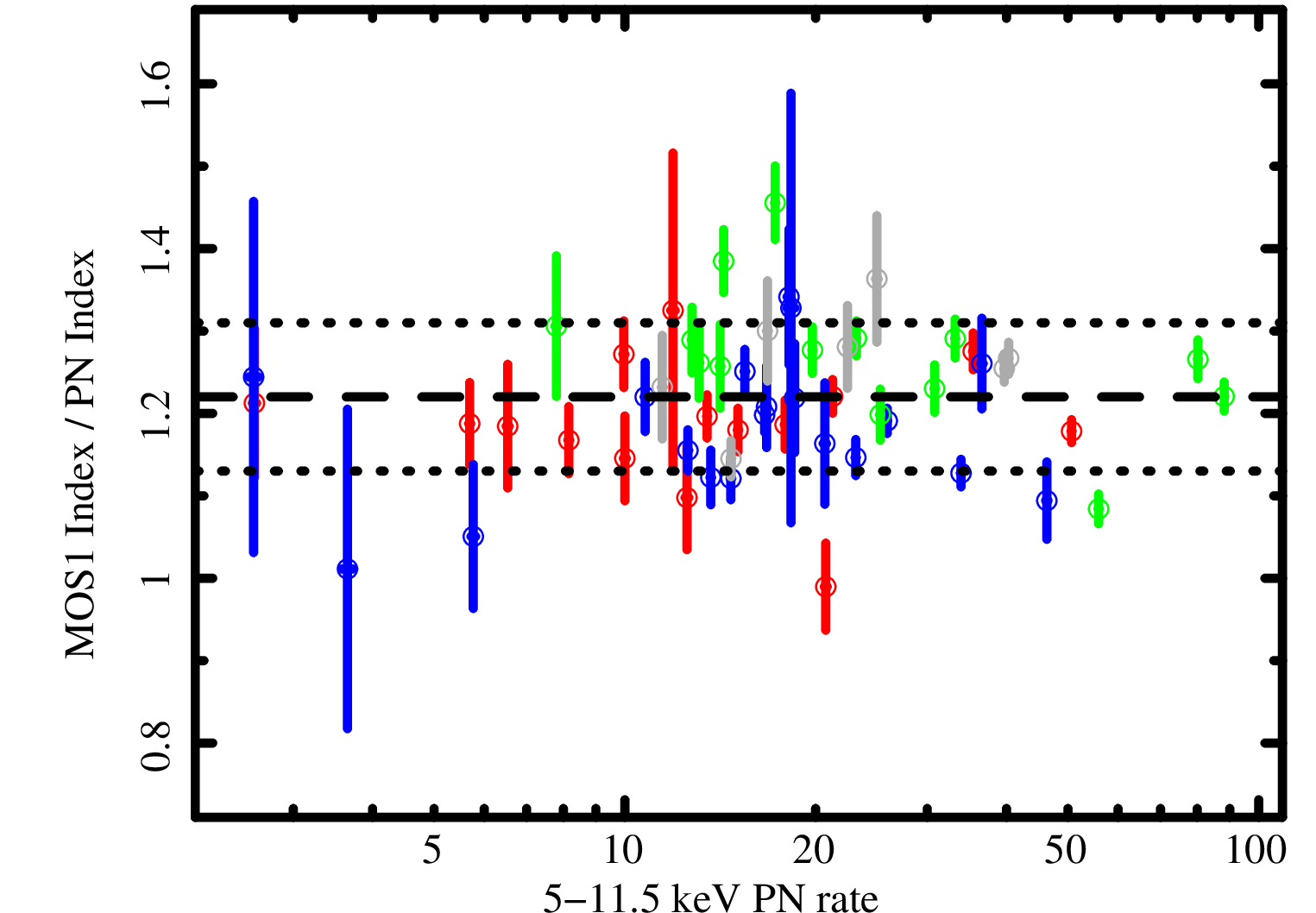}

\hspace{1.0cm}

\includegraphics[width = 5.5cm, angle= -90]{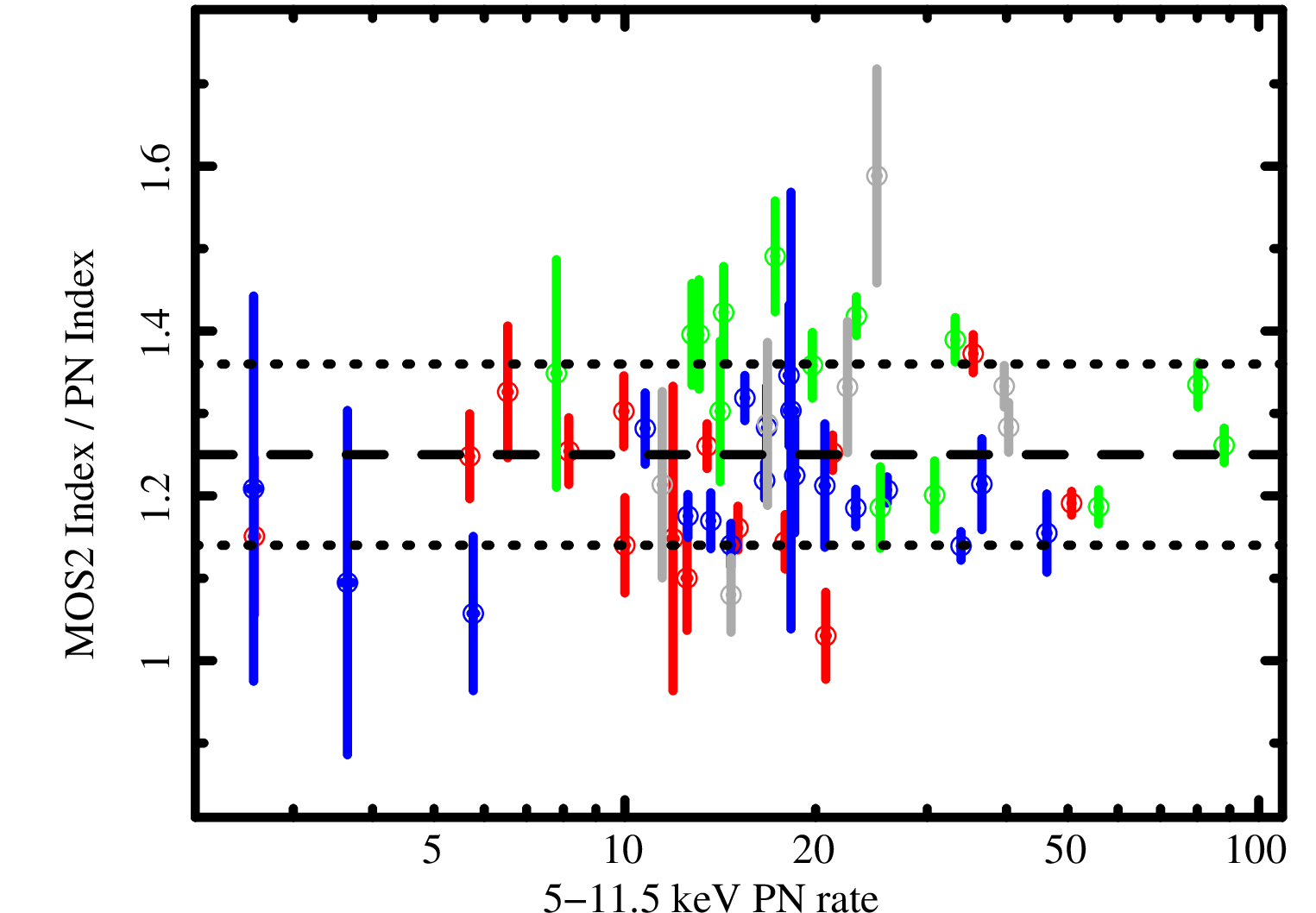}
\caption{Ratio of the spectral indices measured by MOS1 and MOS2 (top panel), MOS1 and PN (middle panel), and MOS2 and PN (bottom panel) as a function of the rates in the 5--11.5 keV band. The dashed black line is the best-fit constant, while the dotted lines indicate the relative error range. Colors indicate the four epochs:
epoch A in red,  epoch B in blue, epoch C in  green, and epoch D in gray. }
\label{fig:mos2-pn_index}
\end{figure}

Figure~\ref{fig:mos2-pn_index} displays, in the three panels, the ratios between the spectral indices measured by MOS1 and MOS2, MOS1 and PN, and MOS2 and PN as a function of the MOS2 and PN rates in the 5--11.5 keV range. 
The averages of the ratios have also been computed for each epoch and the resulting values are listed in Table~\ref{tab:correlation_fit}, where it is evident that no significant variations are detected between epochs.
Considering this, we assume the average values for the entire dataset to be systematic in the spectral indices. They are as follows: 0.97$\pm$0.05 for the spectral index ratio MOS1--MOS2, and 1.22$\pm$0.09 and 1.25$\pm$0.11 for the MOS1 and MOS2 ratio with PN, respectively.
These constants are plotted with a black line in the three panels of Fig.~\ref{fig:mos2-pn_index}, while its error computed as $rms$ is indicated by the dashed lines. 

\begin{figure}
\centering
\includegraphics[width=5.5cm, angle= -90]{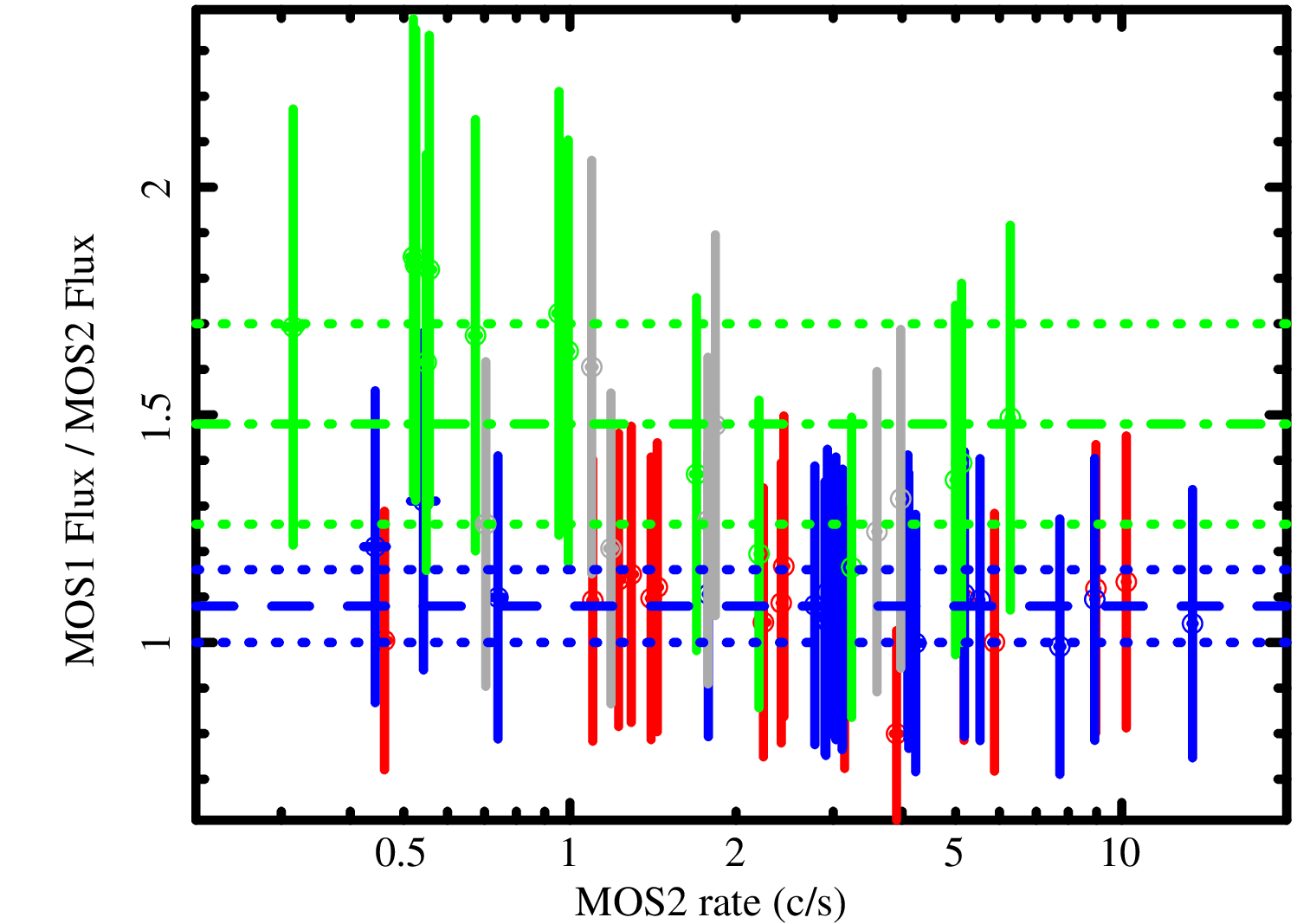}

\hspace{1.0cm}

\includegraphics[width=5.5cm, angle= -90]{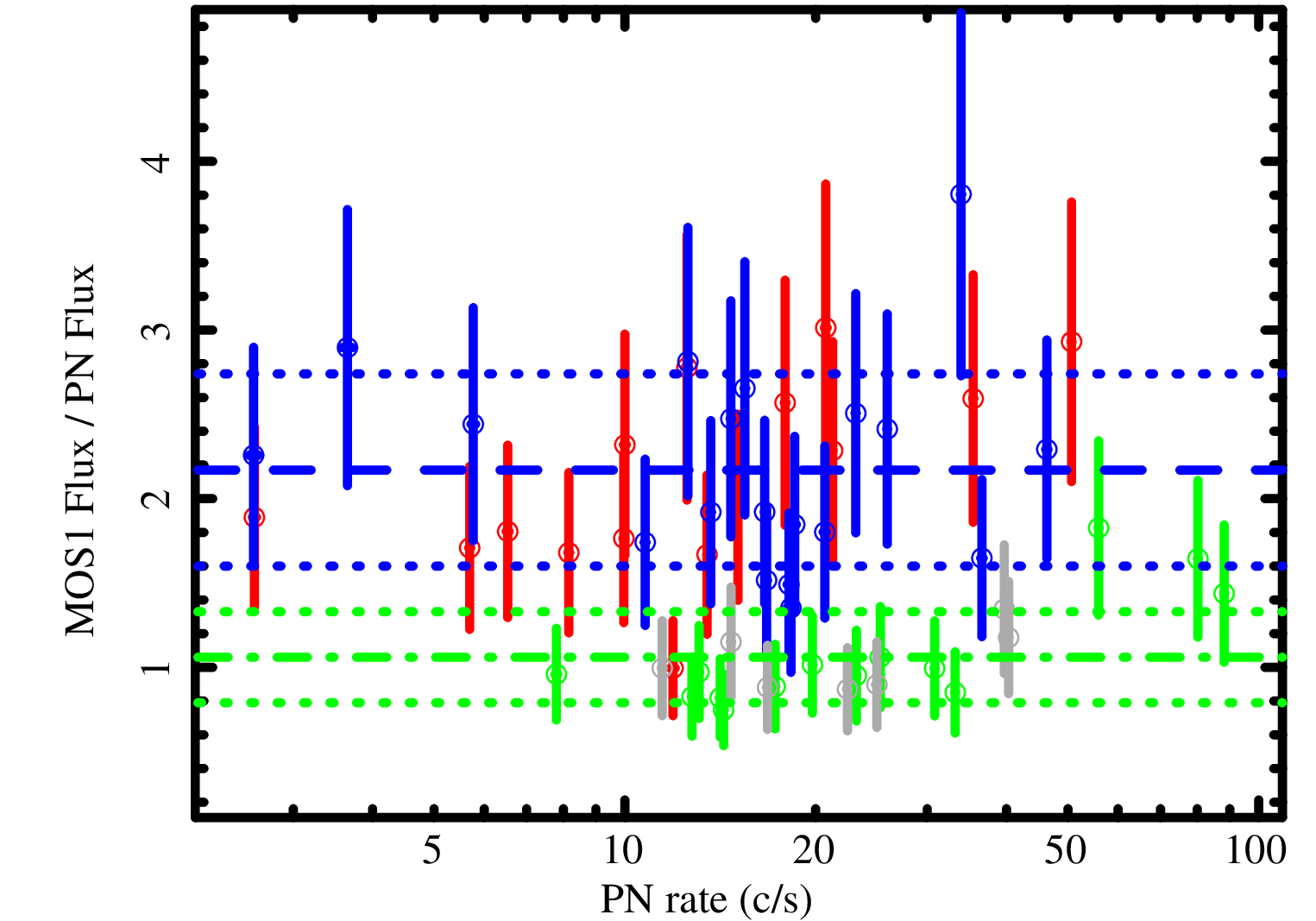}

\hspace{1.0cm}

\includegraphics[width=5.5cm, angle= -90]{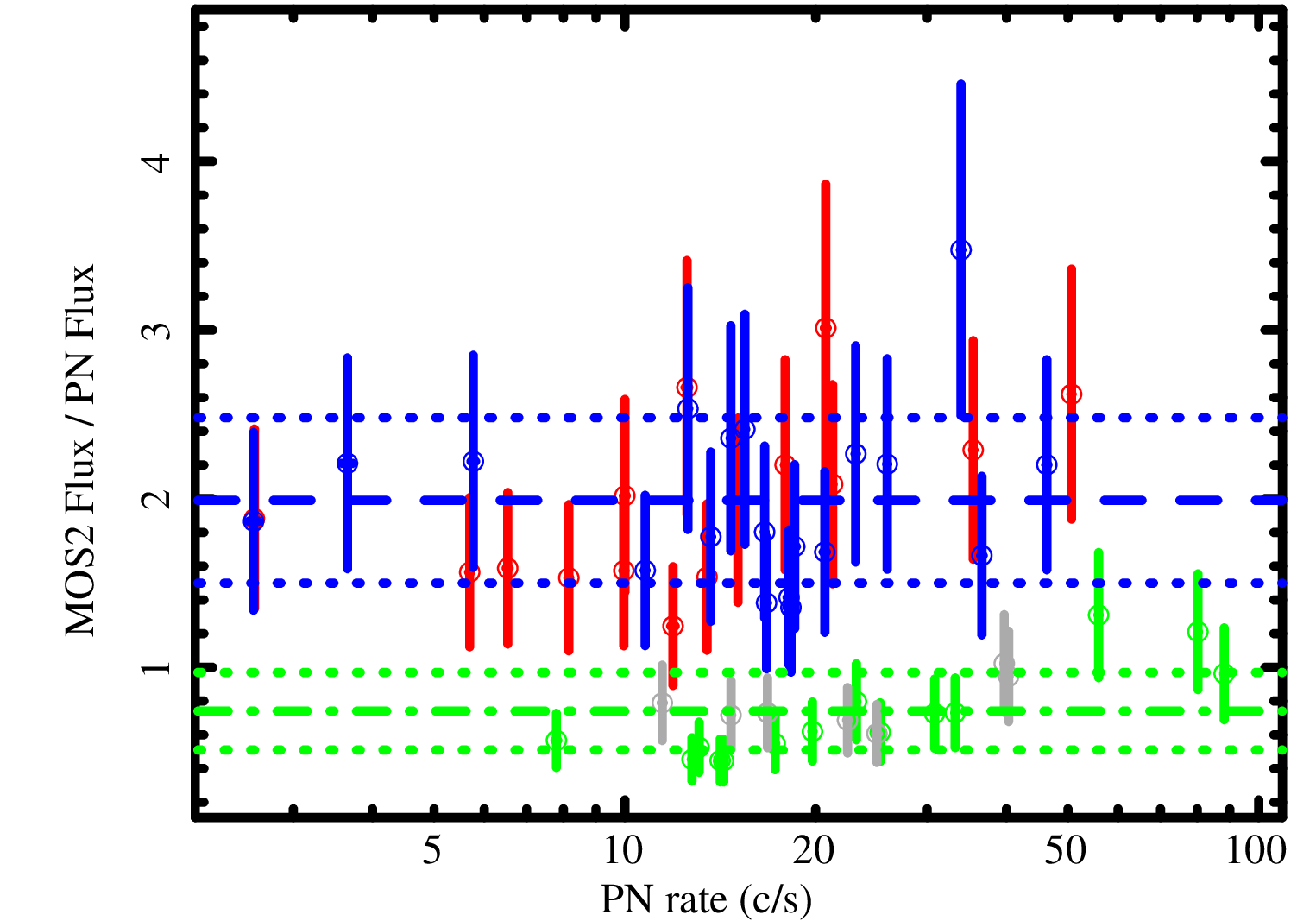}
\caption{Ratio of the 30-300 proton input flux  measured by MOS2 and by PN as a function of the PN rates in the same band. The dashed blue line is the average value for the interval $A$ and $B$, while the dash-dotted green line is the average for $C$ and $D$. The dotted lines indicate the error ranges.
Colors on the points are relative to the four epochs:
epoch A in red,  epoch B in blue, epoch C in  green, and epoch D in gray.   }
\label{fig:mos2-pn_flux}
\end{figure}

The same procedure was applied to quantify the systematic in the fluxes measured by the three instruments. To this purpose, the number of input protons at the telescope pupil in the energy range of 30-300 keV and its errors were computed with the XSPEC command $flux$.  
The plots of the ratios as a function of the rates are shown in the three panels of Fig.~\ref{fig:mos2-pn_flux}. 
Computing the average values in each epoch, we find that, in this case,
consistent fluxes within the statistical uncertainties are measured in epochs $A$ and $B$, and $C$ and $D$. For this reason, two constants have been obtained, one for the year 2001--2002 and one for the 2019--2020. In the first case, the ratio MOS1--MOS2 has an average of 1.08$\pm$0.08, which becomes 1.48$\pm$0.22 for epochs $C$--$D$.
The average values for the ratios MOS1--PN and MOS2--PN are: 2.17$\pm$0.57 and 1.99$\pm$0.49 for $A$ and $B$, while they are 1.06$\pm$0.27 and 0.74$\pm$0.23 for $C$ and $D$, respectively. The two values are shown in Fig.~\ref{fig:mos2-pn_flux} with dashed and dash-dotted lines, respectively.

\begin{table} [th]
\small
\caption{Average value of the ratios between corresponding spectral parameters  detected by the three instruments.}            
\label{tab:correlation_fit}      
\begin{tabular}{l c c c c }     
\hline\hline  
   & \multicolumn{1}{c}{$A$} & $B$ & $C$ & $D$ \\
\hline
\multicolumn{5}{c}{Spectral Index ratio} \\
MOS1/MOS2 &  0.99$\pm$0.06  & 0.98$\pm$0.03 & 0.95$\pm$0.04 & 0.98$\pm$0.06\\
MOS1/PN   & 1.19$\pm$0.07  & 1.18$\pm$0.08 & 1.27$\pm$0.08 & 1.26$\pm$0.06\\
MOS2/PN   &  1.20$\pm$0.08  & 1.21$\pm$0.07 & 1.33$\pm$0.09 & 1.30$\pm$0.14\\
\\
\multicolumn{5}{c}{30--300 keV Flux ratio} \\
MOS1/MOS2 &  1.07$\pm$0.09  & 1.09$\pm$0.07 & 1.56$\pm$0.22 & 1.34$\pm$0.14\\
MOS1/PN   &  2.13$\pm$0.5  & 2.20$\pm$0.59 & 1.07$\pm$0.31 & 1.04$\pm$0.17\\
MOS2/PN   &  1.98$\pm$0.48  & 2.01$\pm$0.50 & 0.71$\pm$0.26 & 0.78$\pm$0.14\\
\hline\hline
\end{tabular}
\tablefoot{Errors are the $rms$ with respect to the average.}
\end{table}

\subsection{Black body}
As was already stated, the black body component is a phenomenological model introduced to take into account the small excess in fitting the whole 2-11.5 keV range with a simple power law. For this model, the best-fit spectral parameters are evaluated with large errors; however, the ratios of the $kT$ and flux values are consistent with those in Table~\ref{tab:correlation_fit} relative to the spectral index and flux of the power law.

\section{Discussion}
\label{sect:discussion}

From the analysis performed on the selected 55 observations of flares, we detected variation in the MOS response to protons with time, identified two components for the spectral model, and evaluated the systematic errors in the best-fit values of both components. The MOS data collected from flares in the epoch 2001--2002 appear to be inconsistent with those from 2019-2020.  MOS1 and MOS2  exhibit statistically equivalent counts a couple of years after launch, while a noticeable reduction occurs after two decades, causing a different count ratio (MOS2 registers lower counts than MOS1). This discrepancy is not reconcilable with the loss of two MOS1-CCDs.
The reduction is proportional across all energy bins, as is evidenced by the lack of dependency of the spectral index ratio as a function of time or count rate (see Fig.~\ref{fig:mos2-pn_index}). 
Considering that the matrices are relative to only single events, we checked for some observations if this reduction could be due to a different distribution of the event grades in 2019-2020 flare photon lists, but no significant differences were obtained.

The effect might be related to the variation in the electrode layer structure observed primarily for MOS2 due to an increase in the surface layers of nitrogen and oxygen and due to the addition of a carbon layer deposited on the electrodes over the course of the mission \citep{Plucinsky2017}. 
From the average rate ratios of the two MOS cameras with respect to PN, as is shown in Table~\ref{tab:rate_ratio}, we assessed the average reduction in the proton effective area between the $C$ and $D$ epochs compared to the $A$ and $B$ epochs, accounting for the loss in MOS1 due to damage in the two CCDs. We found that the MOS1 effective area decreases to 68$\pm$5\% of its value in 2001, while MOS2 has decreased to 38$\pm$6\%. 
Our results are, then, in agreement with studies of the contamination layers that have have formed on the MOS cameras since launch, which indicate higher contamination in the MOS2 camera compared to MOS1, while no contamination was observed on PN \citep{Plucinsky2017}.

In any case, the spectral analysis of MOS flares necessitates a new version of the matrices, with variations mainly involving the computation of the total flux at the telescope input pupil. However, such corrections can only be implemented if the real causes of this phenomenon are understood. If this is not possible, the discrepancy of the matrices with respect to the real transmission can be considered a systematic in the input flux measurements.

Concerning the spectral analysis, the most important information is that a single power law is not able to adequately fit the entire energy range of 2--11.5 keV for most of the spectra, while it is a suitable model in the range of 5--11.5 keV, and its extrapolation to lower energies constitutes the most of the emission.  In fact, the differences in the power law rate with respect to the total detection are $\sim$20\% for the MOS and $\sim$5\% for the PN, and  the 5--11.5 keV power law spectral indices are well correlated  with the hardness ratios (see Fig.~\ref{fig:alfahr_pn}).  
To take into account the soft excess, a second spectral component was included. The most suitable model was a phenomenological black body.

However, we also investigated the possibility of a physical steepening of the power law using two different models: a broken power and a power law with a log parabola spectral index. Neither model gives acceptable results. For the broken power law, the unreliability comes from the impossibility of confining the spectral indices below the energy breaks, probably because of the limited validity range of the matrices. 
The log-parabola power law produces acceptable fits for about half of the spectra, but no coherence between the MOS and PN best values of $\alpha$ and $\beta$ was obtained. 
In summary, with the present version of the proton matrices and from the analysis presented in this paper, it is not possible to state that there is any steepening of the spectrum.
The new version of the matrices or the analysis of a larger sample of spectra could add more items to this point.

\begin{figure}
\centering
\includegraphics[width=6cm, angle= -90]{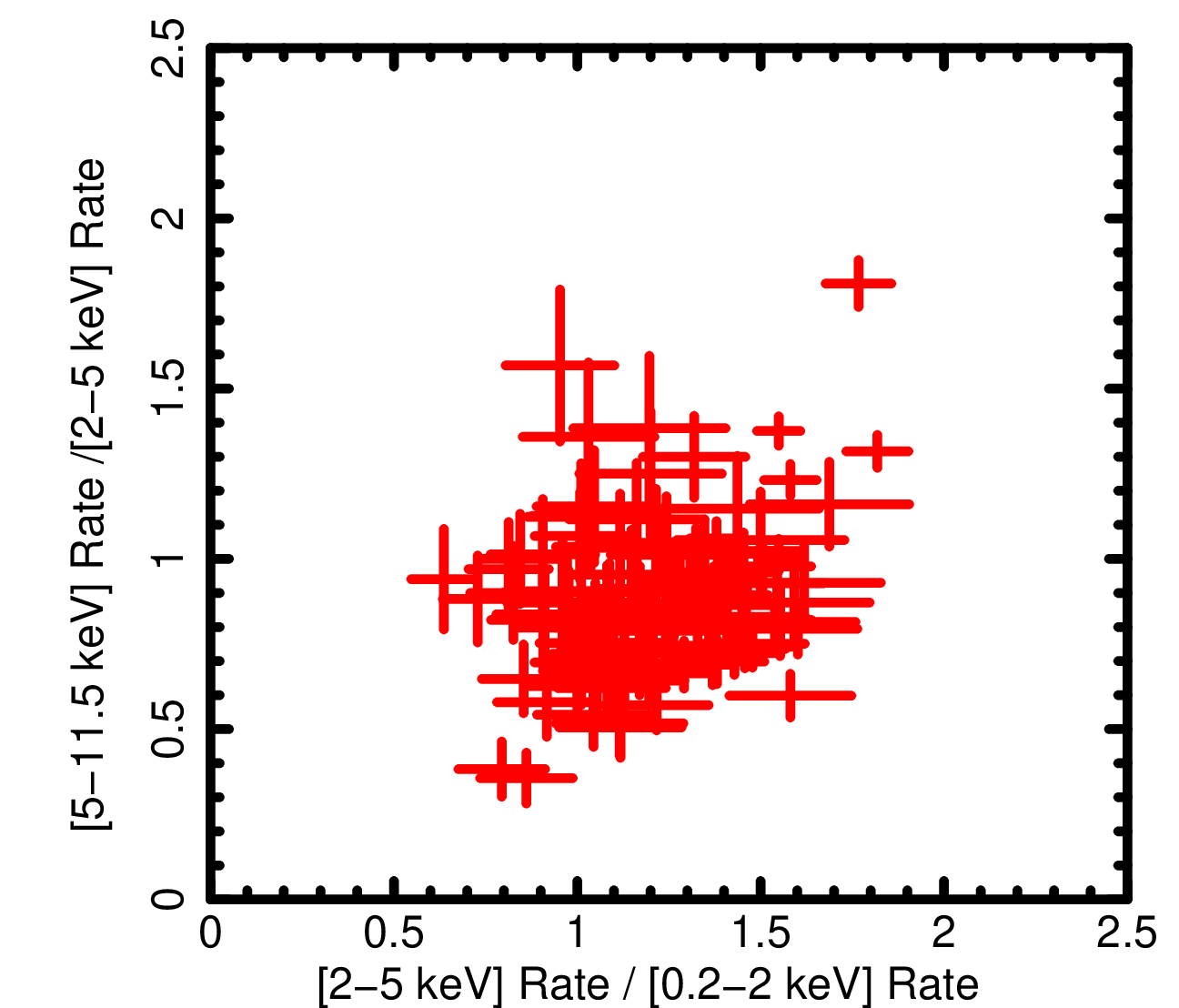}

\vspace{1.0cm}

\includegraphics[width=6cm, angle= -90]{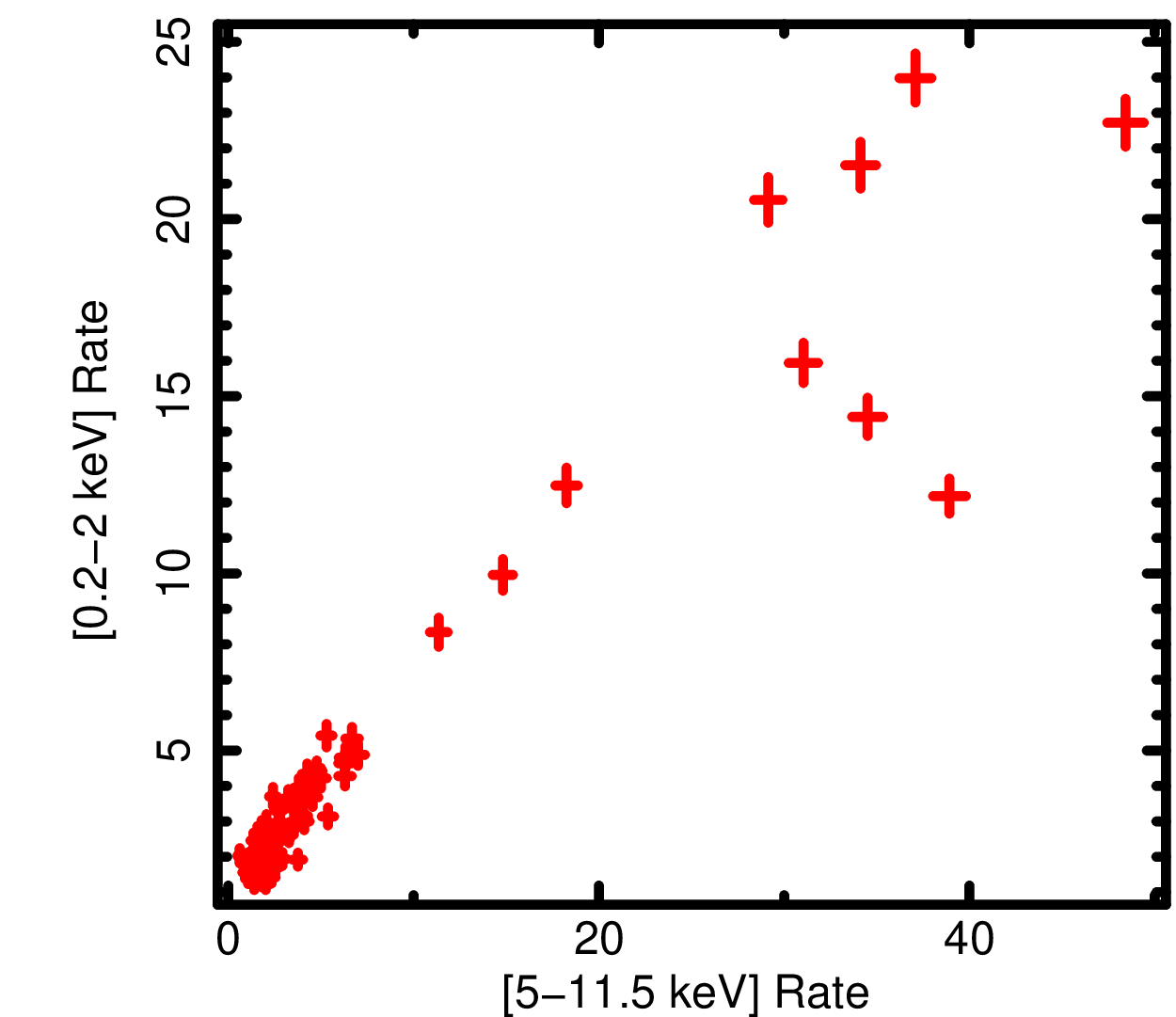}
\caption{Analysis of spectral variability in the observation ID 0037980401:
the top panel shows the color-color diagram and the correlation between rates in the ranges 0.5--2 keV and  5--11.5 keV is presented in the bottom panel. } 
\label{fig:color-color}
\end{figure}

To assess the spectral variability of the black body as a second physical component, we computed a color-color diagram from a few observations (0011830201, 0037980401, and 0052140201). With this aim, we selected light curves in different energy ranges — soft (0.5--2 keV), intermediate  (2--5 keV),  and hard (5--11.5 keV) — and computed the intermediate-soft and hard-intermediate ratios with a time-bin of 50 seconds. Considering that the black body component contributes mostly below 5 keV, we expect that  0.5--2 keV rates could give a strong indication of its variability that we cannot obtain from the spectral analysis being the matrices not valid in this range. As a further check, we also correlated the 0.5--2 keV rates with the 5--11.5 keV ones.  The top panel of Fig.~\ref{fig:color-color} shows as an example the color-color diagram of observation ID 0037980401, with the correlation between rates shown in the bottom panel. 
Throughout the color-color diagram, points are concentrated around an average with variation always lower than 30\% for both the intermediate-soft ratio and the hard-intermediate one. Moreover, a very strong correlation between rates in the ranges of 0.5--2 keV and  5--11.5 keV is always observed. 
We conclude, then, that there is no significant difference in the spectral variability of the two components, and that if this second component is real, it must vary coherently with the power law. However, because we are limited by the validity range of the matrices, a more physical origin — that is, a component coming from heavier ions — cannot be entirely excluded. 

We strongly think that all these results suggest that the power law is the main model for the proton input spectra, and that the second component, necessary to fit the data well at lower energies, is an artifact of the matrices. As is shown in \citetalias{Fioretti2024}, the relative flux of the second component in the MOS analysis increases with the passive material of the optical filters. This, together with the lower excess found in the PN, which is back-illuminated and does not include the electrode structure in front of the depletion region, brings us to the conclusion that it can be attributed to imprecise modeling of the proton transmission at the focal plane in the simulator used for generating the response matrices.
In this case, considering its small contribution to the total count, it can be accounted for by a systematic factor in the input flux derived from the power law.

It is worth pointing out that in some of the PN spectra, the power law overestimates the low energy count rates, and the corresponding MOS spectra have a slightly lower excess with respect to spectra with the same rates. For these, the input proton spectrum is probably flattening below 2 keV, and the simple power law is not applicable. 

This work has also allowed us to quantify the systematic uncertainties in the spectral analysis of flares with the proton matrices. These results are quite simple: the three detectors always measure coherent spectral indices independently from the epoch. The differences are of $\sim$3\% between MOS1 and MOS2 and about 20\% between MOS and PN. This higher discrepancy could be an effect of incorrect modelling of transmission at lower energies, which is more severe in the MOS response matrix.

The measurement of the total flux input at the telescope pupil  gives less clear results because of two different behaviors observed in the data: one in the years 2001-2002 (epochs $A$ and $B$) and the other in 2019-2020 ($C$ and $D$). This makes it difficult to disentangle effects arising from the response matrix from those attributed to variations in the instrument response. The discussion of the total 30-300 keV input flux is then addressed for the 2001--2002 dataset and the 2019--2020 dataset separately. 
\\
In epochs $A$ and $B$, MOS1 and MOS2 exhibit statistically equivalent proton fluxes at the telescope pupil 
(see Fig.~\ref{fig:mos2-pn_flux}).  However, discrepancies arise in the fluxes measured by the MOS and by the PN, revealing a systematic difference of approximately a factor of two (MOS measures twice the PN flux). The systematic increase of 20\% in the MOS spectra indicates that the different excess at lower energies is likely affecting the fitting of the primary power law, with an impact on the input proton fluxes.

In the $C$ and $D$ datasets, the two MOS detectors do not give coherent fluxes, even when accounting for the loss of the two CCDs in MOS1. The systematics of MOS with respect to PN has a more complex behavior. Apart from the second factor identified from 2001--2002 data, an additional factor need to be included. In our context, this factor can be computed from the ratio between the average rates in this period and the average ones in 2001--2002.  However, without understanding the observed rate decrease of the two MOS detectors, it is difficult to correct the response matrix. 

\section{Conclusions}
\label{sect:conclusion}
This paper presents an analysis of 55 flare spectra detected during December-January and July-August of 2001--2002 and 2019--2020, with the main task being to test the matrices on spectra of proton flares and to study the differences in the measured best-fit parameters  between the MOS and PN  proton response matrices.
The spectra have been deconvolved within XSPEC using as a model a power law plus a black body. The main results of this analysis can be summarized in the following points:

\begin{itemize}
\item MOS data show some inconsistency in the detected rates across the 20 years of the satellite's operation with respect to the corresponding PN ones;
\item the physical model representative of the proton spectra at the input of the telescope is a power law, while the black body is a phenomenological component introduced to take into account the low energy excess coming from the extrapolation to lower energies. In any case, this black body component contributes 21\% for the MOS and 5\% for the PN to the total flux in the 2--5 keV energy range; 
\item the systematic error in the spectral indices measured by the two MOS detectors is 3\%, while that of the MOS with respect to the PN is $\sim$24\%;
\item a systematic difference within a factor of two between  MOS and PN input fluxes was determined across the four epochs, as a result of inaccuracies in the proton response matrices and changes in the instrument proton detection across 20 years of operations. This factor was computed directly from  rates, although a more detailed analysis than that presented in this paper is necessary to clarify the causes.
\end{itemize}
A future version of the matrices would include all the elements pointed out by this work and some more specific analysis would be required for us to better understand and model the 2019--2020 loss of transmission efficiency. A systematic analysis of flare events across all the operation life of the satellite could give some clarification.
Moreover, some laboratory measurements on the proton transmission from the filters would allow us to validate Geant4 for this kind of interaction.
To conclude, this work is also necessary for a future X-ray observatory carrying on-board grazing incidence optics, such as Athena,
to produce with the same physical model matrices necessary for predicting the proton flux
at the focal plane.

\begin{acknowledgements} Authors thank the anonymous referee for the comments and suggestions on the paper that really improved it.
TM thanks G. Cusumano and A. D'A\'{i} for their useful comments on the paper. 
The research leading to these results has received funding from European Union’s Horizon 2020 Programme under the AHEAD2020 project (grant agreement n. 871158), the Accordo Attuativo ASI-INAF n. 2019-27-HH.0 and the EXACRAD - CTP ESA contract n. 4000121062/17/NL/LF.
\end{acknowledgements}

%
%
 \bibliographystyle{aa}
\bibliography{sp}
\end{document}